\documentclass{article}

\usepackage{graphicx, verbatim, url, amssymb, amsmath, amsfonts, amsthm, latexsym,epstopdf}
\usepackage{algorithm, caption}
\usepackage[noend]{algorithmic}

\newtheorem{theorem}{Theorem}

\newtheorem{proposition}{Proposition}
\newtheorem{corollary}{Corollary}
 
\algsetup{indent=2em} 

\def\SS{{\cal S}}

\def\vv{{\hat \nu}}

\def\vg{{\bar \nu}}
\def\vr{{\bar \nu}^{R}}

\def\bx{{\bar x}}
\def\by{{\bar y}}
\def\bS{{\bar S}}
\def\bT{{\bar T}}
\def\PP{{\cal P}}
\def\pp{{\bar p}}
\def\qq{{\bar q}}

\title{Vulnerability and power on networks}

\author{Enrico Bozzo \\
Department of Mathematics and Computer Science \\ 
University of Udine \\
\url{enrico.bozzo@uniud.it} \and 
Massimo Franceschet\\
Department of Mathematics and Computer Science \\ 
University of Udine \\
\url{massimo.franceschet@uniud.it} \and
Franca Rinaldi \\
Department of Mathematics and Computer Science \\ 
University of Udine \\
\url{franca.rinaldi@uniud.it}}

\begin{document}
\maketitle

\begin{abstract}
Inspired by socio-political scenarios, like dictatorships, in which a minority of people exercise control over a majority of weakly interconnected individuals, we propose vulnerability and power measures defined on groups of actors of networks. We establish an unexpected connection between network vulnerability and graph regularizability. We use the Shapley value of coalition games to introduce fresh notions of vulnerability and power at node level defined in terms of the corresponding measures at group level. We investigate the computational complexity of computing the  defined measures, both at group and node levels, and provide effective methods to quantify them. Finally we test vulnerability and power on both artificial and real networks.
\end{abstract}

\section{Introduction} \label{introduction}

Our investigation moves from the observation that there exists a recurrent topology in many real-life scenarios characterized by a majority of individuals (that we call the victims), with rare connections among them, that are linked to a minority of people (that we call executioners).  It can be portrayed as a sparse periphery of victims linked to a restricted core of executioners, a sort of generalization of the star topology. In fact, as we will see, the nature of the relationship between victims and executioners may have different semantics depending on the application domain, for instance control or support. 
  
In this paper we conduct a formal investigation of the described topology in the context of network science. We define a vulnerability measure on groups of nodes of an undirected network that quantifies the tendency of a set of actors to be the victims with respect to some smaller group of executioners. We also define a symmetric power measure that assesses the capacity of a group of actors to play the role of executioners with respect to some larger pool of victims. We extend the defined notions of vulnerability and power at the level of network, leading to a characterization of vulnerable networks.  

We discover an unexpected connection between the notion of network vulnerability and that of graph regularizability, a seasoned concept in graph theory. Besides building an interesting bridge between modern network science and traditional graph theory, this result provides us with a method to decide the sign of the vulnerability of a network (positive, null, or negative). We then tackle the  problem of quantifying the exact vulnerability value of a network and finding the set of nodes that determines such vulnerability score. It turns out that, for networks with null or positive vulnerability, this problem can be solved  by exploiting a reduction to the minimum 2-vertex cover problem. We furthermore map the general problem to an integer linear programming model and prove that, whenever the network has non-negative vulnerability, a single continuos relaxation of the model can be exploited to solve the problem. As for networks with negative vulnerability,  we show that the solution of the integer linear programming model can be reduced to the solution of one linear programming problem for each node of the network. 

We then make a detour through game theory. In accordance with a well-established game-theoretic approach to define node centrality in networks, we define a cooperative game over a network in which players are the nodes, coalitions are the groups of nodes, and payoffs of coalitions are defined by the vulnerability (or power) measures on groups of nodes. Hence we interpret the Shapley value of each player in such a game as a centrality measure at node level: the measure represents the average marginal contribution made by each node to the vulnerability (or power) of every coalition of nodes. This allows us to define sophisticated vulnerability and power measures for nodes that take into consideration the corresponding measures for sets of nodes. Notably, we provide closed-form expressions for the Shapley values of both vulnerability and power that can be computed in linear time with respect to the size of the network.

Finally, we test the proposed vulnerability and power measures, at the levels of nodes, sets and network, over artificial networks (random and scale-free graphs) as well as real networks (social and technological networks). We use artificial graphs to investigate the relationship between vulnerability and robustness of networks as defined by algebraic connectivity, as well as for estimating the probability of being a vulnerable network. We use vulnerability and power measures on real networks to reveal meaningful properties of the structure of these networks, as well as to empirically study the correlation between node power and node degree in a network.

The rest of the paper is organized as follows. In Section \ref{informal} we give  two application scenarios for the problems here investigated. Section \ref{formal} does the formal work, defining and investigating vulnerability and power from various angles. The experimental investigation on artificial and real networks is discussed in Section \ref{experiments}. We review the related literature in Section \ref{related} and draw our conclusions in Section \ref{conclusion}. 

\begin{figure}[t]
\begin{center}
\includegraphics[scale=0.40, angle=-90]{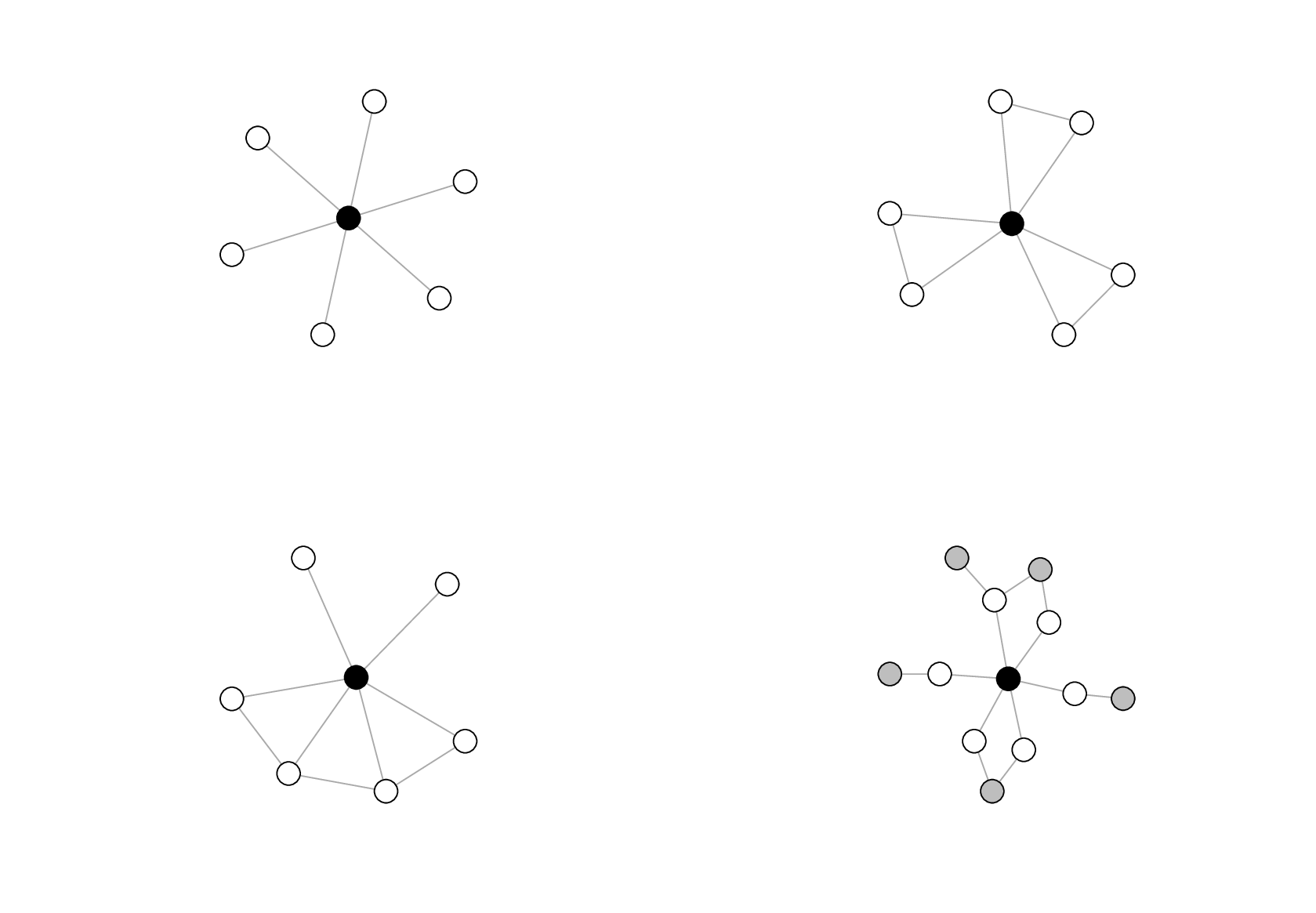}
\end{center}
\caption{Four different network topologies.}
\label{stars}
\end{figure}

\section{Application domains} \label{informal}

In this part we explore two application domains of the notions of vulnerability and power introduced in this paper. The first application domain interprets the relationship between executioners and victims as \textit{control}. Victims are larger in number than executioners, are poorly connected among them, and are controlled by executioners, meaning that there exist no link between a victim and an external actor different from victims and executioners. The result is that executioners can potentially exercise control over victims, since victims can hardly communicate among them and cannot reach external sources. 

This topology is adopted, for instance, in dictatorships. Meetings and associations among people (the victims) are prohibited. Links of victims to external sources of information are hampered. This is accomplished, for instance, by imposing limitations to the use of Internet and popular social networking services. On the other hand, communication necessarily flows only between the dictator or a group of few individuals (the executioners) and the isolated victims. The crucial role of Internet and in particular of social networking services (Twitter in particular) during the uprisings of the Arab Spring has been largely acknowledged. These media have been used by insurgents to break isolation with the external world as well as to organize the internal revolution. These communication links decreased the \textit{vulnerability} of victims with respect to the executioners. 

Further instances of a similar topological exploitation are described in \cite{KISB11}; we quote a couple of historical examples in the following: \textit{
``Plantation owners in Hawaii a century ago expressly hired workers who spoke different native languages to ensure that communication among them would be limited, thus discouraging labor action. And the extraordinary longevity of the Ottoman Empire (1300-1918) and its remarkable integration and taxation of diverse ethnic and religious communities was based on a network structure that made peripheral elites dependent on the center, communicating only with the center rather than with one another.'' 
}

Consider the topologies depicted in Figure \ref{stars}. The archetypal power-vulnerability topology is the star shown in the top-left network: the black node exercises control over a large set of independent white nodes. The set of peripheral victims is vulnerable, and the central executioner is powerful. The  central black node loses much of its control in the top-right configuration: although all white nodes are still connected to it, each white node is also linked to at least another white node. Hence the central black node does not control any white node anymore. The situation depicted in the bottom-left network is intermediate with respect to the previous cases: although the number of bonds between white nodes is the same as in the previous case (3 connections), the distribution of the links penalize the white nodes. Indeed, two of them are still isolated from their white mates and connected only to the black center, which maintain some of its power. Finally, in the bottom-right network, although white nodes are independent, as in the star graph, they are connected to the black node as well as to many other grey nodes. Hence white nodes are not vulnerable and the black node is not powerful.

The second broad application domain is about the influence of social networks on health \cite{BEG00}. A social network is a natural mean to capture and represent social relationships. These relationships  are classified in five categories: social capital, social influence, social undermining, companionship, and social support \cite{HI08}. We are interested in particular in social support that expresses the reciprocal assistance between actors of the social network. Social support is always intended to be helpful, is consciously provided, and if it tries to influence the receiver it is provided in an interpersonal context of caring, trust and respect \cite{HI08}. 
The influence of social support on health have been thoroughly studied; however, few is known on the influence of the topological properties of social networks, such as  diameter, clustering coefficient, degree distribution, and centrality, on social support \cite{CGA10}.

Our view is that vulnerability is a meaningful structural property of a network in relation to social support. More specifically, we argue that networks that are not vulnerable are good models for the exchange of reciprocal assistance. In non-vulnerable networks, each actor can count on the reciprocal help of some neighbor\footnote{A property that we formally show in Proposition \ref{propo:Hall} of Section \ref{formal}.}, a simple idea that is in fact employed by the buddy systems of the Unites States Armed Forces and of the Boy Scouts of America. On the other hand, vulnerable networks contain fragments in which a group of independent actors are connected only to a few central actors; in case of need, most of the independent actors will remain without support. The central actors are good spots for the establishment of a public or professional assistance service.

Consider again the topologies of Figure \ref{stars}. The star topology (top-left) is the worst assistance model: all white actors can receive assistance from only one supporter, the black central actor. Hence all white actors but one are not going to receive any help. This topology identifies, however, the central actor as a perfect spot for a public or professional support server. The bottom-left structure is a somewhat better model of assistance: all white nodes but one can receive support. Indeed, out of the six white actors, four of them can help each other, while a fifth one can receive assistance from the black central actor. On the other hand, the models on the right hand of the picture are good structures for social support. In the bottom-right network, five white actors can receive support from the same number of grey actors, and the last white actor can be assisted from the central black actor. In the top-right topology, all white actors can assist each other, even without the help of the central black actor.

\section{Vulnerability and power on networks} \label{formal}

We start by formally defining the notion of vulnerability. Let $G = (V, E)$ be an  undirected connected graph. For every subset $T \subseteq V$,  we denote by $N(T) =  \{j \in V: \text{there is } i \in T \text{ such that } ij \in E\}$ the set of the neighbors of the nodes in $T$ and by $\SS(G)$ the collection of the independent sets of $G$, i.e., those subsets $S \subseteq V$ such that $N(S) \cap S = \emptyset$. Hence an independent set is a set such that no two vertices in the set are linked by an edge.

We introduce a  {\em vulnerability function} $v_{G}: 2^{V} \rightarrow \mathbb{Z}$  defined by
\begin{equation} \label{eq:vfunction}
v_{G}(T) =  |T| - |N(T)|~~~~~~~~T \subseteq V. 
\end{equation}

Since for every set $T \subseteq V$ each node in $T \cap N(T)$ gives a null contribution to $v_{G}(T)$, the vulnerability function $v_{G}(T)$ can be equivalently expressed as
\begin{equation}
v_{G}(T) = |I(T)| - |N(T)\setminus T| \label{eq:v3bis}
\end{equation}
where $I(T) = T \setminus N(T)$ denotes the independent set containing all the nodes of $T$ that have no neighbor in $T$. One might divide $v_{G}(T)$ by the maximum value it takes (which is $n-2$ on a connected graph), so that the resulting vulnerability lies between $-1$ (minimum vulnerability, corresponding to the vulnerability of the central node of a star network with $n$ nodes) and $1$ (maximum vulnerability, corresponding to the vulnerability of the set of peripheral nodes of a star network with $n$ nodes).

The definition of vulnerability, which is central in this work, claims that a set is vulnerable when it is large and it is connected to few neighbors. Equivalently, a set is vulnerable when it contains a large independent set with few neighbors outside the set. Consider again examples in Figure \ref{stars}. The set $W_1$ of white nodes in the top-left graph $G_1$ is vulnerable: it contains 6 nodes with only 1 neighbor, hence $v_{G_1}(W_1) = 6 - 1 = 5$. Notice that $W_1$ is an independent set, hence $I(W_1) = W_1$. The vulnerability of the white node set $W_2$ in the bottom-left network $G_2$ is largely reduced: the set $W_2$ has 6 members, as before, but the neighbor set $N(W_2)$ contains now 5 nodes, hence $v_{G_2}(W_2) = 6 - 5 = 1$. Notice that $I(W_2)$ is different from $W_2$ and contains 2 nodes, while $N(W_2)\setminus W_2$ contains 1 node. The set $W_3$ of white nodes in the top-right graph $G_3$ in not vulnerable: $v_{G_3}(W_3) = 6 - 7 = -1$. We have moreover that $I(W_3) = \emptyset$. Finally, the set $W_4$ of white nodes in the bottom-right graph $G_4$  in also not vulnerable, but for a different reason. Indeed, $W_4 = I(W_4)$ is independent and contains 6 nodes, the same number of nodes of $N(W_4)$, hence $v_{G_4}(W_4) = 6 - 6 = 0$.     

The {\em vulnerability} $\vg_{G}$ of the network $G$ is the maximum vulnerability of a non-empty independent set of nodes in $G$:
 \begin{equation}
\vg_{G} = \max_{\emptyset \neq S \in \SS(G)} v_{G}(S) \label{eq:barnuG}. 
\end{equation}
We say that $G$ is {\em vulnerable} if ${\vg}_{G} > 0$, i.e., there exists an independent set $S$ such that $|S| > |N(S)|$. On the contrary, in non-vulnerable networks, $|S| \leq |N(S)|$ for every independent set.

A weaker notion of vulnerability can be defined by maximizing the function $v_{G}(T)$ over all the subsets of $V$, not only the independent ones, that is by setting
 \begin{equation}
\vv_{G} = \max_{T \subseteq V} v_{G}(T) \label{eq:nuG}. 
\end{equation}
We define $\vv_{G}$ as {\em weak vulnerability} of the network $G$. Clearly $\vg_G \leq \vv_G$ and,  since $\emptyset \subseteq V$ and $v_{G}(\emptyset) = 0$, then  $\vv_{G} \geq 0$ for each graph $G$. Moreover, the following proposition holds.

\begin{proposition} \label{propo:P1} It holds $\vg_{G} \neq \vv_{G}$ if and only if $\vg_{G} < 0$.
\end{proposition}
\proof Assuming  $\vg_{G} < \vv_{G}$ and, by contradiction, $0 \leq \vg_{G} < \vv_{G}$,  let $\bar T$ be a subset of $V$ such that $v_{G}(\bar T) = \vv_{G}$. Then $v_{G}(\bar T) = |I(\bar T)|  - |N(\bar T)\setminus \bar T| > 0$ and this  implies that the independent set  $I(\bar T)$ is  not empty. From $N(I(\bar T)) \subseteq N(\bar T) \setminus \bar T$ we obtain $v_{G}(I(\bar T)) \geq v_{G}(\bar T)$ and thus $\vg_{G} \geq \vv_{G}$, a contradiction.  The opposite implication follows from the fact that $\vv_G \geq 0$. \qed

From the proof of the above proposition it follows that if $\vv_{G} > 0$ and $\bar T$ is an optimal solution of problem (\ref{eq:nuG}), then also the independent set $I(\bar T)$ is optimal. Moreover, if  $\vv_{G} = 0$, then, since $v_G(\emptyset) = 0$, the empty set, which is an independent set, is an optimal solution of (\ref{eq:nuG}). It follows that we can write:
\begin{equation} \label{eq:idvv}
\vv_{G} = \max_{S \in \SS(G)} v_{G}(S).
\end{equation}

\subsection{Determining if a network is vulnerable} \label{vulnerability}

As a first aspect, we consider the problem of determining if a network $G$ is vulnerable or not. In graph theory the networks $G$ with $\vg_{G} \leq 0$ and $\vg_{G} < 0$ have been characterized    from several perspectives. A first characterization arises from the study of quasi-regularizable and regularizable graphs. We recall that a graph $G$ is \textit{quasi-regularizable} if it is possible to assign non-negative integer weights to the edges of the graph in such a way that the sum of the weights over the edges incident in any node is the same non-null value. The graph is called \textit{regularizable} if these weights can be chosen strictly positive. An alternative characterization, useful from a computational point  of view, involves the notion of $2$-matching. 
A {\em 2-matching} is an assignment  of weights 0, 1 or 2 to
the edges of the graph with the property that the sum of weights of the edges incident in any
node is at most 2. If this sum is exactly 2 for each node,  the $2$-matching  is called {\em perfect}. The notion of 2-matching someway generalizes the  notion of {\em matching}. We remind that a matching $M$ is a subset of edges with the property that different edges of $M$ cannot have a common endpoint. A matching $M$ is called perfect if every node of the graph is the endpoint of (exactly) one edge of $M$. In the following we will exploit the fact that 2-matchings are strictly related to 2-vertex covers, where a {\em 2-vertex cover} is an assignment of weights 0, 1 and 2 to the nodes such that for each edge the sum of the weights of its endpoints is at least 2. In turn, the notion of 2-vertex cover someway generalizes the notion of vertex cover. We remind that a vertex cover $A$ is a subset of nodes with the property that each edge of the graph has at least one endpoint in $A$.

We summarize the main relations between the above concepts and the properties $\vg_{G} \leq 0$ and  $\vg_{G} < 0$ in the following two theorems.

\begin{theorem} \label{quasiregular} Let $G = (V, E)$ be a connected undirected graph. Then the following conditions are equivalent:

\begin{enumerate}
\item $|S| \leq |N(S)| $ for every independent set $S \subseteq V$, i.e., $\vg_{G} \leq 0$;
\item $G$ is quasi-regularizable \cite{Berge1981};
\item $G$ admits a perfect 2-matching \cite{Tutte1953}.
\end{enumerate}

\end{theorem}

\begin{theorem}  \label{theo:regular} Let $G = (V, E)$ be a connected undirected graph. Then the following conditions are equivalent:
\begin{enumerate}
\item  $|S| < |N(S)| $ for every independent set $\emptyset \neq S \subseteq V$, i.e., $\vg_{G} < 0$;
\item $G$ is a regularizable graph that is not elementary bipartite, where a bipartite graph is elementary if every edge is contained in a perfect matching \cite{Berge1978a};
\item $G$ is a 2-bicritical graph, i.e., for each node $i \in V$ the graph $G(V \setminus \{i\})$ admits a perfect 2-matching \cite{Pulleyblank}.
\end{enumerate}
\end{theorem}

We will see in Section \ref{computation} how the problem of determining if a graph admits a perfect 2-matching can be solved in polynomial time by finding a maximum matching on a bipartite graph. Therefore Theorems \ref{quasiregular} and \ref{theo:regular}  imply that one  can determine in polynomial time the sign of the vulnerability $\vg_{G}$ of a graph. 

The following proposition, that follows from Hall's Theorem \cite{LP1986}, points out an interesting property of non-vulnerable networks: each node of any independent set can be matched with a different neighbor. 

\begin{proposition} \label{propo:Hall} Let $G$ be a network with $\vg_G \leq 0$. Then for each $S \in {\cal S}(G)$, $S \neq \emptyset$,  there exists
an injective map $\phi: S \rightarrow N(S)$ such that $\phi(i) \in N(\{i\})$ for each $i \in S$. 
\end{proposition}

\subsection{Computing the vulnerability of a network} \label{computation}

In this section we present two polynomial methods to compute the vulnerability of a network. The first method is a strongly polynomial algorithm and works for non-regularizable networks.  The second method, valid for the general case, is based on an integer linear programming model of the problem. We show that the solution of this model  can actually be reduced to the solution of $|V|$ linear programming  problems, one for each node of the network.

A polynomial method to compute the vulnerability of non-regularizable graphs, i.e., graphs $G$ with  $\vg_G \geq 0$, is provided by the theory of the 2-matchings and 2-vertex covers.  For the sake of completeness, we report here the main results that justify the method and refer the reader to \cite{LP1986} for a complete exposition of the subject.


In the following, the sum of  the components of a vector $z$ is called the {\em size} of $z$ and is denoted by $|z|$. In graph theory, the minimum size of a 2-vertex cover of a graph $G$ is denoted by  $\tau_{2}(G)$ and the maximum size of a 2-matching is denoted by $\nu_{2}(G)$. It is well known that the maximum possible size of a 2-matching is $|V|$ and that a 2-matching is perfect if and only if it has size $|V|$.

The following two results state an important relationship between the weak vulnerability $\vv_{G}$ of a graph, the maximum size of a 2-matching and the minimum size of a 2-vertex cover. 
\begin{theorem} \label{theo:2-cover} If $G = (V, E)$ is an undirected graph, then 
 \begin{equation} \label{eq:2-cover}
\nu_{2}(G) = \tau_{2}(G) =  \min_{S \in {\cal S}(G)} |V| - |S| + |N(S)| = |V| - \vv_G.
\end{equation}
\end{theorem}
\proof For the two relevant equalities $\nu_{2}(G) = \tau_{2}(G) =  \min_{S \in {\cal S}(G)} |V| - |S| + |N(S)|$ we refer to \cite{LP1986}. The last equality  directly follows from identity (\ref{eq:idvv}). \qed
 
Given a 2-vertex cover $\bar u$ of minimum size an independent set $\bar S$ with $v_{G}(\bS) = \vv_{G}$ is given by
\begin{equation} \label{eq:sol}
\bar S = \{i \in V: \bar u_{i} = 0\}.
\end{equation}
Note that, since $\bar u_{i} + \bar u_{j} \geq 2$ for each $ij \in E$, the set $\bar S$ is in fact an independent set of $G$ and $\bar u_{j} = 2$ for each $j \in N(\bar S)$. Moreover, the optimality of $\bar u$ implies  $\bar u_{k} = 1$ for each $k \in V \setminus (\bar S \cup N(\bar S))$, so that $|\bar u|  = 2 |N(\bS)| + |V| - |\bS| - |N(\bS)| = |V| - |\bar S|  +  |N(\bar S)|$. In particular, $\bar S = \emptyset$ if and only if $\bar u_{i} = 1$ for each $i \in V$ and thus $|\bar u| = |V|$ and $\vv_{G} = 0$. As a consequence $\bar S$ can be the empty set only if $\vg_G \leq 0$ and it is necessarily the empty set if $\vg_G < 0$.

Theorem \ref{theo:2-cover} and Proposition \ref{propo:P1} immediately imply the following corollary.

\begin{corollary} \label{coro:2-cover} If $\vg_{G} \geq 0$,  then $\vg_{G} = |V| - \nu_{2}(G) = |V| - \tau_{2}(G)$.
\end{corollary}

Based on the previous results, the following theorem gives the complexity of solving problem (\ref{eq:barnuG}) for non-regularizable graphs.

\begin{theorem} \label{theo:computation}  Let $G = (V, E)$ be an undirected connected graph. The problem of determining  a non-empty independent set of maximum vulnerability  $\vg_G$ can be solved in time $O(|V|^{\frac{1}{2}}|E|)$ if $\vg_G > 0$,  and in time $O(|V|^{\frac{3}{2}}|E|)$ if $\vg_G = 0$. In particular, the sign of $\vg_{G}$ can be determined  in time $O(|V|^{\frac{3}{2}}|E|)$.
\end{theorem} 
\proof As it follows from  Theorem \ref{theo:2-cover} and Corollary \ref{coro:2-cover}, if $\vg_{G} \geq 0$ then $\vg_{G} = |V| - |\bar u|$ where $\bar u$ is any 2-vertex cover of minimum size 
of $G$. As shown in \cite{LP1986}, the problem of finding a 2-vertex cover $\bar u$ of minimum size reduces to that of finding a minimum vertex cover on a bipartite graph  with $2|V|$ nodes and $2|E|$ edges. Now, as reported in  \cite{Schrijver2003}, the minimum vertex cover problem on bipartite graphs can be solved in $O(|n|^{\frac{1}{2}}|m|)$ where $n$ is the number of nodes of the graph and $m$ the number of edges. 
Given a 2-vertex cover $\bar u$ of minimum size, let $\bS$ be
the independent set defined in (\ref{eq:sol}). If $\bS \neq \emptyset$, as it always happens when $\vg_{G} > 0$, then $\bS$ is an optimal solution of problem (\ref{eq:barnuG}). Otherwise, if $\bS = \emptyset$, then  $\vv_{G} = 0$ and $G$ is quasi-regularizable. In this case, by item 3 of Theorem \ref{theo:regular}, $G$ is non-regularizable if and only if for at least one node $k \in V$ the graph $G(V \setminus \{k\})$ does not admit a perfect 2-matching. By  Theorem \ref{theo:2-cover} this is equivalent to both  $\nu_{2}(G(V \setminus \{k\})) = \tau_{2}(G(V \setminus \{k\})) < |V| - 1$ and $\vg_{G(V \setminus \{k\})}) > 0$. Therefore, if $\vg_{G} = 0$, such a node $k$ can be found by solving at most $|V|$ instances of the 2-vertex cover problem of minimum size, 
one for each node of the graph, with a global time requirement $O(|V|^{\frac{3}{2}}|E|)$. If $\bS$ is an independent set of maximum vulnerability in the graph $G(V \setminus \{k\})$, then it must be $v_{G(V \setminus \{k\})}(\bS) = 1$, $k \in N_{G}(\bar S)$ and $v_{G} (\bar S) = 0$. So $\bS$ is an optimal solution of problem (\ref{eq:barnuG}). On the contrary, when $\vg_{G} < 0$, the procedure returns $\vv_{G(V \setminus \{k\})} = 0$ for each $k \in V$. \qed

  We remark that the problem of computing the sign of the vulnerability $\vg_{G}$ of a graph (without finding an independent set of maximum vulnerability) can be tackled by solving a minimum size 2-matching problem (at most $|V|$ minimum size 2-matching problems if $\vv_{G} = 0$) instead of a 2-vertex cover problem. This does not change the complexity of the procedure since the last two problems have not only the same optimal value, as stated in Theorem \ref{theo:2-cover}, but their solving algorithms share a common main part \cite{LP1986}.

The computation of $\vg_{G}$  further simplifies when  $G$ is a bipartite graph. 

\begin{corollary} If $G = (V_{1} \cup V_{2}, E)$ is a bipartite graph, then a non-empty independent set of maximum vulnerability $\vg_G$ can be found in $O(|V|^{\frac{1}{2}}|E|)$ by solving a maximum matching problem on $G$. 
\end{corollary}
\proof Being $V_{1}$ and $V_{2}$ independent sets of $G$ and  $N(V_{1}) = V_{2}$, $N(V_{2}) = V_{1}$, then  either $v_{G}(V_{1}) \geq 0$ or $v_{G}(V_{2}) \geq 0$.  Thus ${\bar v}_G \geq 0$ and Corollary \ref{coro:2-cover} applies. Now, as shown in \cite{LP1986}, a 2-matching of maximum size in a bipartite graph can be obtained by simply assigning weight 2 to the edges of a maximum matching. The statement follows from the fact that a maximum matching in $G$ can be found in $O(|V|^{\frac{1}{2}}|E|)$ \cite{Schrijver2003}. In particular, if $\vg_{G} = 0$, then both $V_{1}$ and $V_{2}$ are independent sets of maximum vulnerability. \qed

When $\vg_G < 0$ the equivalence between the problems of maximizing the vulnerability function over the non-empty sets of $\SS(G)$ and that of finding a 2-vertex cover of minimum size does not hold anymore.  In order to solve problem (\ref{eq:barnuG}) in the general case we adopt an integer linear programming approach. A 0-1 linear programming model of the problem can be defined by introducing  two binary variables $x_{i}$ and $y_{i}$ for each $i \in V$  with the meaning that $x_{i} = 1$ if $i \in S$, 0 otherwise, and $y_{i} = 1$ if $i \in N(S)$, 0 otherwise. The model is
\begin{eqnarray}
{\cal P}_{G}: ~~~~~~~~~~\max ~~~\sum_{i \in V} ~(x_{i} &-& y_{i}) \nonumber\\
x_{i} + x_{j} & \leq & 1 ~~~~~~ij~~\in~~E \label{eq:stable}\\
y_{j} & \geq & x_{i} ~~~~ij~~\in~~E \label{eq:ns1}\\
y_{i} & \geq & x_{j} ~~~~ij~~\in~~E \label{eq:ns2}\\
\sum_{i \in V} x_{i} & \geq & 1 \label{eq:ns3}\\
x_{i}, y_{i} & \geq	 & 0 ~~~~~~~i \in V. \label{eq:bounds}\\
x_{i} & \in & \mathbb{Z}~~~~~~~i \in V. \label{eq:integer}
\end{eqnarray}
Constraints (\ref{eq:stable}) assure that the set $S$ of the nodes $i$ with $x_{i} = 1$ is an independent set, constraints (\ref{eq:ns1}) and (\ref{eq:ns2}) force to 1 all the variables $y_{j}$ associated with nodes in $N(S)$, while constraint (\ref{eq:ns3}) excludes the solution corresponding to $S = \emptyset$.  Note that we have omitted the constraints $x_{i}, y_{i} \leq 1$ and the integrality constraints on the $y$ variables since they are anyway satisfied in every optimal solution.

Our next task is that to show that problem ${\cal P}_{G}$ can actually be solved by solving $|V|$ linear programming problems. To this aim for each node $k \in V$ consider the integer linear programming problem ${\cal P}_{G}(k)$ obtained from problem ${\cal P}_{G}$ by substituting constraint (\ref{eq:ns3}) with the constraint $x_{k} = 1$, that is by forcing node $k$ to belong to an optimal solution, and denote by $\vg_{G}(k)$ its optimal value. Moreover, denote by ${\cal P}_{G}^{R}(k)$ the continuos relaxation of problem ${\cal P}_{G}(k)$ and by $\vg^{R}_{G}(k)$ its optimal value. 
The next result states that every problem ${\cal P}_{G}(k)$ can be solved by solving its relaxation ${\cal P}_{G}^{R}(k)$.

\begin{figure}[t]
\begin{center}
\includegraphics[scale=0.40]{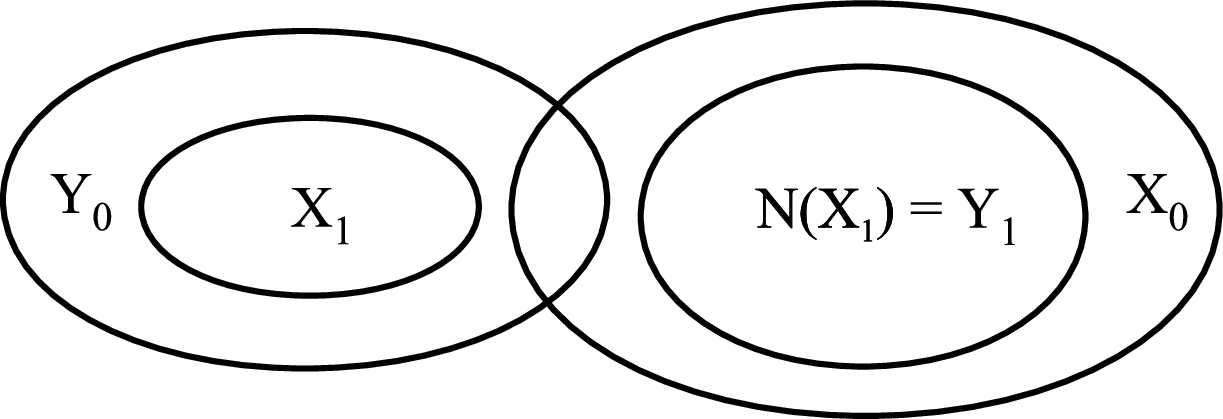}
\end{center}
\caption{Inclusions among the sets $X_{0}$, $X_{1}$, $Y_{0}$, $Y_{1}$ and $N(X_{1})$ used in the proof of Theorem \ref{theorem:relax2}.}
\label{teo5a}
\end{figure}

\begin{figure}[t]
\begin{center}
\includegraphics[scale=0.40]{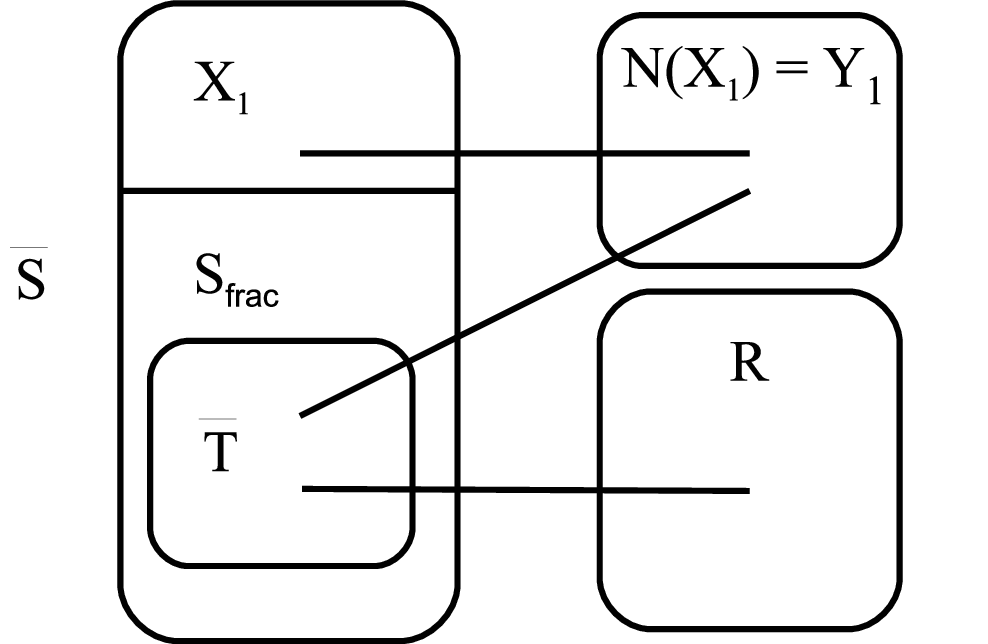}
\end{center}
\caption{Relations among the sets $\bar S$, $S_{frac}$, $X_{1}$, $N(X_{1})$ and sets  $\bT$ and  $R$  used in the proof of Theorem \ref{theorem:relax2}.}
\label{teo5b}
\end{figure}

\begin{theorem}  \label{theorem:relax2} Let $G=(V, E)$ be an undirected graph. Then for each $k \in V$ it holds $\vg_{G}(k)  = \vr_{G}(k)$ and an optimal solution of problem ${\cal P}_{G}(k)$ can be derived by any optimal solution of problem ${\cal P}_{G}^{R}(k)$. \\
\end{theorem}
\proof Let $(\bx, \by)$ be an optimal solution of problem $\PP_{G}^{R}(k)$. For  $r \in \{0,1\}$ define $X_{r} = \{ i \in V: \bx_{i} = r\}$ and $Y_{r} = \{ i \in V: \by_{i} = r\}$. Consider the sets  $X_{1}$ and $N(X_{1})$. The set $X_{1}$, containing node $k$, is not empty. Moreover, by constraints (\ref{eq:stable}), (\ref{eq:ns1}) and (\ref{eq:ns2}) for each $j \in N(X_{1})$ it holds $\bx_{j} = 0$ and $\by_{j} = 1$, thus $N(X_{1}) \subseteq X_{0} \cap Y_{1}$. Moreover the optimality of $(\bx, \by)$ implies $\by_{i} = \max_{j \in N(\{i\})} \;\bx_{j} $ for each $j \in V$ and thus, in particular, $X_{1} \subseteq Y_{0}$ and $Y_{1} = N(X_{1})$.  The relations among the sets $X_{0}$, $X_{1}$, $Y_{0}$, $Y_{1}$ and $N(X_{1})$ are shown in Figure \ref{teo5a}. By the above considerations $\bx_{i} - \by_{i} = 1$ for each $i \in X_{1}$ and the set $X_{1}$ is contained in the set $\bS = \{ i \in V: \bx_{i} > \by_{i}\}$. From the constraints (\ref{eq:ns1}) and  (\ref{eq:ns2}) it also follows that
\begin{equation}
\label{eq:vr=0}
\quad \quad \quad \quad \by_{j} \geq \bx_{i} > \by_{i} \geq \bx_{j} \quad \quad \mbox{ for every }~~i \in \bS, \;j \in N(\{i\}).
\end{equation}
In particular $\bS$ is an independent set of $G$ and, since $\bx_{j} - \by_{j} \leq 0$ for each $j \in V \setminus \bS$, it holds 
\begin{equation} \label{eq:frac0bis}
\vr_{G} (k) = \sum_{i \in V} (\bx_{i} - \by_{i}) \leq \sum_{i \in \bS} (\bx_{i} - \by_{i}) + \sum_{j \in N(\bS)} (\bx_{j} - \by_{j}).
\end{equation}
In order to prove the statement it is now sufficient to show that the right-hand side of (\ref{eq:frac0bis}) is not greater than $|X_{1}| - |N(X_{1})|$, since this implies that the integer solution corresponding to the independent set $X_{1}$ defines an optimal solution of problem  ${\cal P}^{R}_{G}(k)$ and thus an optimal solution of problem  ${\cal P}_{G}(k)$. The thesis holds when $\bS = X_{1}$ since in this case the right-hand side of (\ref{eq:frac0bis}) is equal to $|X_{1}| - |N(X_{1})|$. Let us assume, on the contrary, that the set $S_{frac} = \bS \setminus X_{1}$ is not empty and rewrite (\ref{eq:frac0bis}) as
\begin{equation} \label{eq:fracbis}
\vr_{G}(k) = \sum_{i \in V} (\bx_{i} - \by_{i}) \leq |X_{1}| - |N(X_{1})| + \sum_{i \in S_{frac}} (\bx_{i} - \by_{i}) + \sum_{j \in N(S_{frac}) \setminus N(X_{1})} (\bx_{i} - \by_{i}).
\end{equation}
In order to prove that $\sum_{i \in S_{frac}} (\bx_{i} - \by_{i}) + \sum_{j \in N(S_{frac}) \setminus N(X_{1})} (\bx_{i} - \by_{i}) \leq 0$, let us first show that  it holds $|T| \leq |N(T) \setminus N(X_{1})|$ for every $T \subseteq S_{frac}$.  Assume by contradiction that there exists  $\bT \subseteq S_{frac}$ such that $|\bT| > |N(\bT) \setminus N(X_{1})|$ and choose such a set $\bT$ of minimum cardinality. Define $R = N(\bT) \setminus N(X_{1})$. By the above considerations, it holds $R = N(\bT) \setminus Y_{1}$. The relations among the sets $\bar S$,  $S_{frac}$, $X_{1}$ and $Y_{1} = N(X_{1})$ and the sets $\bT$ and  $R$ are shown in Figure \ref{teo5b}.  For $\delta > 0$ sufficiently small the solution $(x', y')$ defined by
\begin{eqnarray}
 x'_{i} =  \bx_{i} + \delta \quad\quad ~i \in \bT \quad \quad &\quad \mbox{and}  \quad &  y'_{i}  =  \by_{i}  - \delta \quad\quad j \in \bT \setminus Y_{0}\\
 x'_{j} =  \bx_{j} - \delta \quad \quad  j \in R \setminus X_{0} &\quad \mbox{and}  \quad &  y'_{j}  =   \by_{j}  + \delta \quad \quad  j \in R\\
 x'_{k} =  \bx_{k}  &\quad \mbox{and}  \quad & y'_{k} =  \by_{k} \quad \mbox{otherwise}
 \end{eqnarray}
is  feasible for ${\cal P}_{G}^{R}(k)$ and its value differs from $\vr_{G}(k)$ by the amount
$$\Delta = \delta \, \big(|\bT|  + |\bT \setminus Y_{0}| - |R \setminus X_{0}| - |R \setminus Y_{1}| \, \big) \geq \delta \,  \big( 2 |\bT|  - |\bT \cap Y_{0}| - 2 |R| + |R \cap X_{0}|  \big).$$
Since constraints (\ref{eq:ns1}) and (\ref{eq:ns2}) imply $N(Y_{0}) \subseteq X_{0}$ we have that $N(\bT \cap Y_{0}) \setminus N(X_{1}) \subseteq R \cap X_{0}$. Thus in the case $\bT \subseteq Y_{0}$ it holds $R \subseteq X_{0}$ and we obtain $\Delta \geq \delta(|\bT| - |R|)$. Otherwise the minimality of $|\bT|$  implies $|\bT \cap Y_{0}|  \leq |N(\bT \cap Y_{0}) \setminus N(X_{1})| \leq |R \cap X_{0}|$ and we obtain $\Delta \geq 2 \delta (|\bT| - |R|)$. Being $|\bT| > |R|$ by assumption,  in both cases we get  $\Delta > 0$  in contradiction with the optimality of $(\bx, \by)$. So we can assume  $|T| \leq |N(T) \setminus N(X_{1})|$ for every $T \subseteq S_{frac}$. By Hall's Theorem \cite{LP1986}, this implies that there exists an injective map $\phi: S_{frac} \rightarrow N(S_{frac}) \setminus N(X_{1})$ such that $\phi(i) \in N(\{i\})$ for each $i \in S_{frac}$. Since property (\ref{eq:vr=0}) implies $\by_{\phi(i)} - \bx_{\phi(i)} \geq \bx_{i} -  \by_{i}$ for each $i \in S_{frac}$, from (\ref{eq:fracbis})  we finally obtain, as required,
$$\vr_{G}(k) = \sum_{i \in V} (\bx_{i} - \by_{i}) \leq |X_{1}| - |N(X_{1})| + \sum_{i \in S_{frac}} (\bx_{i} - \by_{i} + \bx_{\phi(i)} - \by_{\phi(i)}) \leq |X_{1}| - |N(X_{1})|.$$ 
 \qed 

We remark that an argument similar to that used in the proof of Theorem \ref{theorem:relax2} allows to prove that  when $\vg_{G} > 0$ an optimal solution of problem ${\cal P}_{G}$ can be obtained simply by solving its continuous relaxation. 
 
\begin{corollary} The vulnerability $\vg_{G}$ of every undirected network $G = (V, E)$ can be computed in polynomial time. 
\end{corollary}
\proof Since linear programming problems are polynomial \cite{K1980}, the statement follows from Theorem \ref{theorem:relax2} and the fact that $\vg_{G} = \max_{k \in V} \vg_{G}(k)$. \qed

It is worth noticing that for every maximal independent set $S$ of a graph $G$ it holds $N(S) =  V \setminus S$ and hence $v_{G}(S) = |S| - |V \setminus S| = 2|S| - |V|$. It follows that the problem of finding a \textit{maximal} independent set of maximum vulnerability corresponds to the problem of finding an independent set of maximum cardinality, which is known to be NP-hard.

We conclude this section by showing some topological properties of the vulnerability function $\nu_{G}(T)$. We first show that the vulnerability function $\nu_{G}(T)$ is \textit{non-monotonic}. Recall that a  real function  $f$ defined on the collection $2^{V}$  of all the subsets of $V$ is monotonically increasing (respectively, decreasing) if for all $S, T \subseteq V$ with $S \subseteq T$, it holds that $f(S) \leq f(T)$ (respectively, $f(S) \geq f(T)$). Indeed, consider a set $T \subseteq V$ and a node $i \notin T$. Suppose there are $k \geq 0$ neighbors of $i$ not belonging to the neighbors of $T$, that is, $|N(\{i\}) \setminus N(T)| = k$. Then 
$$ \nu_{G}(T \cup \{i\}) = 
   |T \cup \{i\}| - |N(T \cup \{i\})| = 
   |T| + 1 - |N(T)| - k = 
   \nu_{G}(T) + 1 - k  
$$

Hence, if $k = 0$, then  $\nu_{G}(T \cup \{i\}) > \nu_{G}(T)$; if $k = 1$, then $\nu_{G}(T \cup \{i\}) = \nu_{G}(T)$; and if $k \geq 2$, then $\nu_{G}(T \cup \{i\}) < \nu_{G}(T)$.

On the other hand, the vulnerability function $\nu_{G}(T)$ is {\em supermodular}. A real function  $f$ defined on $2^{V}$ is supermodular if for all $S, T \subseteq V$ it holds that $f(S \cup T) + f(S \cap T) \geq f(S) + f(T)$. Moreover, $f$ is called submodular if $g = -f$ is supermodular and $f$ is called modular if $f$ is both supermodular and submodular.
  
\begin{theorem} \label{theo:supermodular} The vulnerability function $v_{G}(T)$ is supermodular. 
\end{theorem}
\proof Since $|T|$ is a modular function it is sufficient to show that $|N(T)|$ is a submodular function. This immediately follows from the fact that for each pair of subsets $S, T \subseteq V$ it holds $|N(S \cup T)| = |N(S)| + |N(T)| - |N(S) \cap N(T)|$  and $N(S \cap T) \subseteq N(S) \cap N(T)$. \qed

We remark that the problem of maximizing an integer-valued supermodular function $f$, i.e., to find a subset $T \subseteq V$ of maximum value $f(T)$, can be solved in strongly polynomial time if $f$ is given by a value giving oracle and the function is bounded \cite{GLS}. So every polynomial algorithm for the maximization of a supermodular function offers, according to Proposition \ref{propo:P1}, an alternative  way to compute the vulnerability $\vg_{G}$ of a vulnerable network. The complexity of these methods \cite{Iwata2008} is, however,  largely dominated by the above described approach based on 2-vertex covers and 2-matchings. 



\subsection{A symmetric perspective: power} \label{sec:power}

Assuming a symmetric perspective, in this section we study two power functions that measure the capacity of a set of nodes to completely control a set of other nodes. To this aim for every $T \subseteq V$ we denote by $B(T) = \{i \in V: N(\{i\}) \subseteq T\}$ the subset of nodes whose neighbors are contained in $T$. By definition, the subset $S(T) = B(T) \setminus T$ is an independent set.   

We define two {\em power functions} $p, q:  2^{V} \rightarrow \mathbb{Z}$  by setting, for each $T \subseteq V$:

\begin{equation}
p_{G}(T) =  |B(T)| - |T| \label{eq:pfunction}
\end{equation}
and  
\begin{equation}
q_{G}(T) =  |S(T)| - |T| \label{eq:barpfunction}
\end{equation}

Hence, a set $T$ is powerful if it is small and controls a large set $B(T)$. Notice that nodes in $B(T)$ do not have connections outside $T$, hence are potentially at the mercy of nodes in $T$. Moreover, nodes in $S(T)$ are controlled nodes that are not themselves controllers. Let us consider again Figure \ref{stars}. The black node $i_1$ in the top-left graph $G_1$ is powerful: it controls all 6 white nodes. We have that $p_{G_1}(\{i_1\}) = q_{G_1}(\{i_1\}) = 6 - 1 = 5$. The power of the black node $i_2$ in the bottom-left graph $G_2$ is severely reduced: it now controls only two nodes, hence $p_{G_2}(\{i_2\}) = q_{G_2}(\{i_2\}) = 2 - 1 = 1$. Graph $G_2$ is useful to distinguish the two power functions. Consider the set $T$ containing the four connected white nodes plus the black node. We have that $B(T)$ is the set of all white nodes, while $S(T) = B(T) \setminus T$ contains only the two white nodes that are not connected among themselves. Hence $p_{G_2}(T) = |B(T)| - |T| = 6 - 5 = 1$ and  $q_{G_2}(T) = |S(T)| - |T| = 2 - 5 = -3$.  The black node $i_3$ in the top-right graph $G_3$ has completely lost its power: it does not control any node, hence $p_{G_3}(\{i_3\}) = q_{G_3}(\{i_3\}) = 0 - 1 = -1$.
Notice that, for all graphs analyzed so far, the power of the black node corresponds to the vulnerability of the complementary set of white nodes (that we computed above), a property that we  formally show in the first item of the next Proposition \ref{relation}. 
Finally, the black node of the bottom-right graph does not control any node, hence its power is $-1$. In this case, because of the grey vertices, the set of white nodes is not the complement of the set containing the only black node.

Power at the graph level is defined as follows:

\begin{equation} \label{ppg}
\pp_G = \max_{ T \subseteq V} p_{G}(T)
\end{equation}
and 
\begin{equation} \label{qqg}
\qq_G = \max_{T \subseteq V: \;S(T) \neq \emptyset} q_{G}(T).
\end{equation}

Since $S(T) \subseteq B(T)$ for each $T \subseteq V$, it holds $\qq_G \leq \pp_G$.
The next proposition points out the strong relationship between $\pp_G$ and $\qq_G$ and the vulnerability notions $\vg_G$ and $\vv_G$ introduced in the previous section. 

\begin{proposition} \label{relation} For every network $G$ it holds that:
\begin{enumerate}
\item $p_{G}(T) = v_{G}(V \setminus T)$ for each $T \subseteq V$;
\item $ \pp_G = \vv_G $ and  $ \qq_G = \vg_G $.
\end{enumerate}
\end{proposition}
\proof  Item 1 follows from the fact that for each $T \subseteq V$ it holds that $B(T) = V \setminus N(V \setminus T)$ and thus
$$p_{G}(T) = |V \setminus N(V \setminus T)| - |T| = |V| - |N(V \setminus T)| - |T| = |V \setminus T| - |N(V \setminus T)| = v_{G}(V \setminus T).$$

We now show item 2 of the proposition. The first identity immediately follows from item 1. About the second identity, we note that for every non-empty independent set  $U$,  it holds that  $U \subseteq S(N(U))$. So we obtain
$$v_{G}(U) = |U| - |N(U)| \leq |S(N(U))| - |N(U)| = q_{G}(N(U))$$
that implies ${\bar v}_G \leq \qq_G$. On the other hand for each $T \subseteq V$ with $S(T) \neq \emptyset$ it holds $N(S(T)) \subseteq T$ and this implies $v_{G}(S(T)) \geq q_{G}(T)$. As a consequence  $\vg_G \geq \qq_G$. \qed

As a consequence of the above result, the problems (\ref{eq:nuG}) and  (\ref{ppg}) are  equivalent. In particular $\bar T$ is an optimal solution of  problem (\ref{eq:nuG}) if and only if $V \setminus \bar T$ is an optimal solution of problem (\ref{ppg}). In the same way,  the problems (\ref{eq:barnuG}) and (\ref{qqg})  are equivalent. In particular if $\bar S$ is an optimal solution of problem (\ref{eq:barnuG}) then  $N(\bS)$ is an optimal solution of problem (\ref{qqg}); conversely, if $\bar T$ is an optimal solution of problem (\ref{qqg}) then $S(\bar T)$ is an optimal solution of problem (\ref{eq:barnuG}).  Moreover, by Proposition \ref{propo:P1}, if $G$ has a non-negative vulnerability $\vg_{G}$ then  $\qq_G = \vg_G = \vv_G = 
\pp_G \geq 0$ and by Theorem \ref{theo:computation} a set of maximum power can be found in polynomial time. Also, item 1 of Proposition \ref{relation}  implies that the power function $p_{G}(T)$, as the vulnerability function $v_{G}(T)$, is non-monotonic and supermodular.
Differently, the power function $\qq_G(T)$ is not supermodular. For instance, for every graph $G$ and each node $i$ it holds $q_G(V\setminus \{i\}) + q_G(\{i\}) = 1 - (|V| - 1) + |S(\{i\})| - 1 > - |V| = q_G(V)$.

\subsection{A game-theoretic definition of power and vulnerability} \label{gamepower}

Both  the power and the vulnerability functions introduced above associate values with subset of nodes, and not with single nodes as it is common for the centrality measures proposed in network theory. In this respect they are, according to the terminology introduced in \cite{EB99}, {\em group centrality measures}. In this section we show how to derive vulnerability and power at node level using a game-theoretic approach. This can be done by using the power and vulnerability functions to define suitable coalitional games on the node set of the network and by considering a classical game solution, the Shapley value. For the game theory notions in this section the reader is referred, among others, to \cite{OR1994}.

In game theory, a characteristic function is commonly used to assign to each coalition  of players a value corresponding to the power of the coalition, i.e., how much these players can globally get if they decide to play together, independently on the other players' actions. A common task in game theory is that of deriving, on the base of the characteristic function,  an assignment of scores to the players as an index of the power of the single players in the game. Probably the most popular and used solution proposed for coalitional games is the {\em Shapley value}. This solution  associates with each game ${\cal G} = (N, w)$, where $N$ is the set of players and $w: 2^{N} \rightarrow \mathbb{R}$ is the characteristic function,  a vector $\phi \in \mathbb{R}^{|N|}$ whose components are given by
\begin{equation} \label{shapley1}
\phi_{i} = \frac{1}{|N|!} \sum_{L \in \Pi} (w(T_L(i) \cup \{i\}) - w(T_L(i)))  \quad \quad i \in N,
\end{equation}
where $\Pi$ denotes the set of all the orders (permutations) of the players and $T_L(i)$, $L \in \Pi$, denotes the coalition formed by the players that precede $i$ in $L$. In other words $T_L(i) = \{k \in N: L(k) < L(i)\}$ where $L(k)$ is the position of node $k$ in the order $L$. 
According to this definition, the score assigned to each player $i$ is the average over all the orders $L$ of the player set $N$ of the contribution that player $i$ gives when it reaches the coalition $T_{L}(i)$. Alternatively, the Shapley value can be expressed in the more compact form 
\begin{equation} \label{shapley2}
\phi_{i} = \sum_{T \subseteq N: i \notin T} \frac{|T|!(|N| - |T| - 1)!}{|N|!} (w(T \cup \{i\}) - w(T))  \quad \quad i \in N.
\end{equation}

The computation of the Shapley value for coalitional games requires, in general, exponential time. As a consequence, despite its interest, this value can be computed using formula (\ref{shapley1}) or (\ref{shapley2}) only for games with a number of players relatively small. Nevertheless, in some cases the particular structure of the characteristic function allows for an explicit formula of the Shapley value of the game.  This favorable situation actually occurs for the power and vulnerability functions we have considered. 

The next theorem gives an explicit expression of the Shapley value for the games defined by the power functions $p_{G}(T)$ and $q_{G}(T)$. The argument used in the proof is similar to the one used in \cite{MASRJ13} for other group centrality measures. 

\begin{theorem} Given a graph $G$, the Shapley values $\phi^{p}$ and $\phi^{q}$ of the coalitional games $(V, p_{G}(T))$ and $(V, q_{G}(T))$ have the expression
\begin{eqnarray}
\phi^{p}_i & = & -1 + \sum_{j \in N(\{i\})} \; \frac{1}{d_j}  \quad~~~~~~~~~i \in V \label{sp}\\
\phi^{q}_i & = & - 1 - \frac{1}{1 + d_i} + \sum_{j \in N(\{i\})} \; \frac{1}{(1 + d_j) d_j } ~~~~~~i \in V \label{sq}
\end{eqnarray} 
where $d_i$ is the degree of node $i$. 
\label{th:power}
\end{theorem}
\proof Let $i$ be a node of $G$.  Given an order $L \in \Pi$, the marginal contributions of $i$ to the set $T = T_L(i)$ with respect to the characteristic functions $p_{G}(T)$ and $q_{G}(T)$, respectively, are\begin{eqnarray}
p_{G}(T \cup \{i\}) - p_{G}(T) & = & |B(T \cup \{i\}) \setminus B(T)| - 1 \label{deltap}\\
q_{G}(T \cup \{i\}) - q_{G}(T) & = & |S(T \cup \{i\}) \setminus S(T)| - |S(T) \cap \{i\}| - 1 \label{deltaq}.
\end{eqnarray} 
It holds that
\begin{eqnarray*}
B(T \cup \{i\}) \setminus B(T) & = & \{ j \in N(\{i\}): N(j) \setminus \{i\} \subseteq T\}\\
S(T \cup \{i\}) \setminus S(T) & = & \{ j \in N(\{i\} \setminus T: N(j) \setminus \{i\} \subseteq T\}.
\end{eqnarray*} 
As a consequence, the only nodes that can give a non-trivial contribution to (\ref{deltap}) and  (\ref{deltaq}) are those in $N(\{i\})$ and possibly, in the case of (\ref{deltaq}), the node $i$. Moreover a node $j \in N(\{i\})$ gives a contribution  to $|B(T \cup \{i\}) \setminus B(T)|$ in expression (\ref{deltap}) only for those orders $L$ where all the nodes in $N(\{j\}) \setminus \{i\}$ belong to $T$, i.e., precede $i$ in $L$. It is easy to verify that number of such orders is  
$$ \binom{|V|}{d_j} (d_j - 1)! (|V| - d_j)! = \frac{|V|!}{d_j}.$$
Similarly, a node $j \in N(\{i\})$ gives a contribution  to $|S(T \cup \{i\}) \setminus S(T)|$ in expression (\ref{deltaq}) only for those orders $L$ where all the nodes in $N(\{j\}) \setminus \{i\}$ precede $i$ and $L(j) > L(i)$. It is easy to verify that the number of such orders is  
$$ \binom{|V|}{d_j + 1} (d_j - 1)! (|V| - d_j -1)! = \frac{|V|!}{d_j(1 + d_j)}.$$
Finally, the orders in which node $i$ gives a  contribution  to  $|S(T) \cap \{i\}|$ in (\ref{deltaq}) are those in which $N(\{i\}) \subseteq T$. The number of these orders is 
$$ \binom{|V|}{d_i + 1} (d_i)! (|V| - d_i - 1)! = \frac{|V|!}{1 + d_i}.$$
Now the expressions (\ref{sp}) and (\ref{sq}) follow immediately from the definition (\ref{shapley1}) of the Shapley value. \qed

We can justify the above result as follows. It states that power rewards actors having a large number of low-degree neighbors. The difference between the two power functions $\phi^{p}$ and $\phi^{q}$ is that the latter, because of the quadratic dependency on the degree of neighbors, is less sensitive to neighbors of relatively high degree. Now, consider a generic node set $T$ and a node $i$ not belonging to $T$. Theorem \ref{th:power} states that the marginal contribution given by $i$ to the power of $T$ is high if $i$ has many neighbors with low degree. Indeed, if $j$ is a low-degree neighbor of $i$, the probability that all neighbors of $j$ are in $T \cup \{i\}$, hence that $j$ is a new victim of $T \cup \{i\}$, is high. On the other hand, if $j$ has many neighbors, then it is unlikely that all of them belong to $T \cup \{i\}$, hence that $j$ is controllable by $T \cup \{i\}$. It follows that a node $i$ that provides the highest increment to the power of a generic set $T$ is a node with many neighbors of unitary degree, that is, node $i$ is the center of a star subgraph. In this case, all the neighbors of $i$ become, for sure, new victims of $T \cup \{i\}$.  On the other hand, a node $i$ that provides the lowest increment to the power of $T$ is a node with no neighbors; in fact it decreases the power of one unity. 

As an example, consider for the umpteenth time Figure \ref{stars}. In all four networks, the black node has the same number of neighbors (the six white nodes). However, these neighbors have different degrees, and this determines different powers for the black vertex. Let us consider, for the sake of simplicity, power $\phi^{p}$. The maximum power, equal to $-1 + 6 = 5$, is achieved by the black node of the star network in the top-left part of the figure. The black node of the bottom-left network has a lower power equal to $- 1 + (1+1+\frac{1}{2}+\frac{1}{2}+\frac{1}{3}+\frac{1}{3}) = -1 + \frac{11}{3} = \frac{8}{3} $. The power of the black node of the top-right network is still lower:  $-1 + \frac{1}{2} \cdot 6 = 2$, and the black node of the bottom-right network has the lowest power equal to $- 1 + (\frac{1}{2} \cdot 5 + \frac{1}{3}) = - 1 + \frac{17}{6} = \frac{11}{6}$. Notice that, if we call $i$ the black node, it always holds that the Shapley-based power $\phi^{p}_{i}$ of $i$ is larger than or equal to the node set power $p_G(\{i\})$ of the singleton $\{i\}$ (that we computed above), a property that we formally show in Proposition \ref{core}.

The thesis that power is in the hands of those connected to powerless actors might be surprising at first sight. Classical recursive centrality measures, like eigenvector and PageRank centrality \cite{F11-CACM}, remunerate those actors that are connected to powerful ones. Nevertheless, the notion has its logic, as sagaciously observed by  \cite{B87}: \textit{``However, in bargaining situations, it is advantageous to be connected to those who have few options; power comes from being connected to those who are powerless. Being connected to powerful others who have many potential trading partners reduces one's bargaining power''}. Bonacich observes in a subsequent footnote that this notion of power appears already in Caplow's and Gamson's well-known theories of coalition formation of late sixties. A related notion of power in a hierarchically structured population of economic agents has been proposed by \cite{BG94}.

Finally, it is worth pointing out that both power measures $\phi^{p}$ and $\phi^{q}$ can be computed in linear time in the size of the graph, that is, in $O(|V| + |E|)$.  

Let us now consider the coalitional game ${\cal G}(V, v_{G})$ defined by the vulnerability function $v_{G}(T)$.
The following proposition shows how the symmetry between the vulnerability and power functions reflects in the symmetry of the Shapley values of the corresponding games. 

\begin{proposition} \label{propo:vulgame}
For every network $G = (V, E)$, the Shapley values $\phi^{p}$ and $\phi^{v}$ of the games  ${\cal G}(V, p_{G})$ and ${\cal G}(V, v_{G})$ are symmetric, i.e.,  $\phi^{v} = - \phi^{p}$.
\end{proposition}
\proof By item 1 of Proposition \ref{relation}, for each $T \subseteq V$ and $i \notin T$ 
$$v_{G}(T \cup \{i\})  - v_{G}(T)= p_{G}(V \setminus (T \cup \{i\})) -  p_{G}(V \setminus T) = - (p_{G}(V \setminus T) - p_{G}(V \setminus (T \cup \{i\}))).$$
Since the contributions of the node $i$ with respect to the sets $T$ and $V\setminus (T \cup \{i\}))$ have the same coefficient in the expression ({\ref{shapley2}) of the Shapley value the statement holds.  
\qed

Games defined by supermodular characteristic functions, as the games defined by the power function $p_{G}$ and the vulnerability function $v_{G}$, are  commonly called {\em convex games} and exhibit some important properties \cite{Shapley71}. One of these properties is that the Shapley value of a convex game ${\cal G} = (N, v)$ always belongs to the {\em core} of the game, i.e.,  the set of the payoffs $a \in \mathbb{R}^{|N|}$ that satisfy the condition $\sum_{i \in S} a_{i} \geq v(S)$ for each coalition $S \subseteq N$. Payoffs in the core are considered robust solutions of the game, since they give to any coalition at least what the coalition can get by itself. In particular, the core of every convex game is not empty.


For completeness we report here a direct proof that the Shapley values $\phi^{p}$ and $\phi^{v}$ belong to the core of the corresponding games.

\begin{proposition} The Shapley values $\phi^{p}$ and $\phi^{v}$ of the games ${\cal G}(V, p_{G})$ and ${\cal G}(V, v_{G})$ belong to the respective cores. \label{core} 
\end{proposition}
\proof In order to show that $\phi^{p}$ belongs to the core of ${\cal G}(V, p_{G})$ it is sufficient to show that for each coalition $T \subseteq V$ it holds $\sum_{i \in T} \sum_{j \in N(\{i\})} \frac{1}{d_{j}} \geq |B(T)|$. Now each node $k \in B(T)$ contributes with a term $\frac{1}{d_{k}}$ to exactly $ |N(\{k\})| = d_{k}$ terms of the left hand side. As a consequence 
$$\sum_{i \in T} \sum_{j \in N(\{i\})} \frac{1}{d_{j}} \geq  \sum_{k \in B(T)} \frac{d_{k}}{d_{k}}  = |B(T)|.$$
Consider the Shapley value $\phi^{v}$ of game ${\cal G}(V, v_{G})$. Propositions \ref{relation} and \ref{propo:vulgame} and the just proved item for $\phi^{p}$ imply that, for each $T \subseteq V$ 
$$v_{G}(T) = p_{G}(V \setminus T)  \leq \sum_{i \in V \setminus T} \phi^{p}_{i} = - \sum_{i \in V \setminus T} \phi^{v}_{i} = \sum_{i \in T} \phi^{v}_{i} $$
where the last identity follows from the fact that, by the efficiency axiom of the Shapley value, $\sum_{i \in V} \phi^{v}_{i} = v_{G}(V) = 0$.   \qed


\section{Experimental analysis} \label{experiments}

In this section we discuss the outcomes of the experiments that we conducted on artificial as well as real networks. We mostly used the computing environment R, and in particular the network analysis package \textit{igraph}. We solved the integer linear programming model for the computation of vulnerability $\vg_{G}$ proposed in Section \ref{formal} using the solver CPLEX 11.2.  

\begin{figure}[t]
\begin{center}
\includegraphics[scale=0.40, angle=-90]{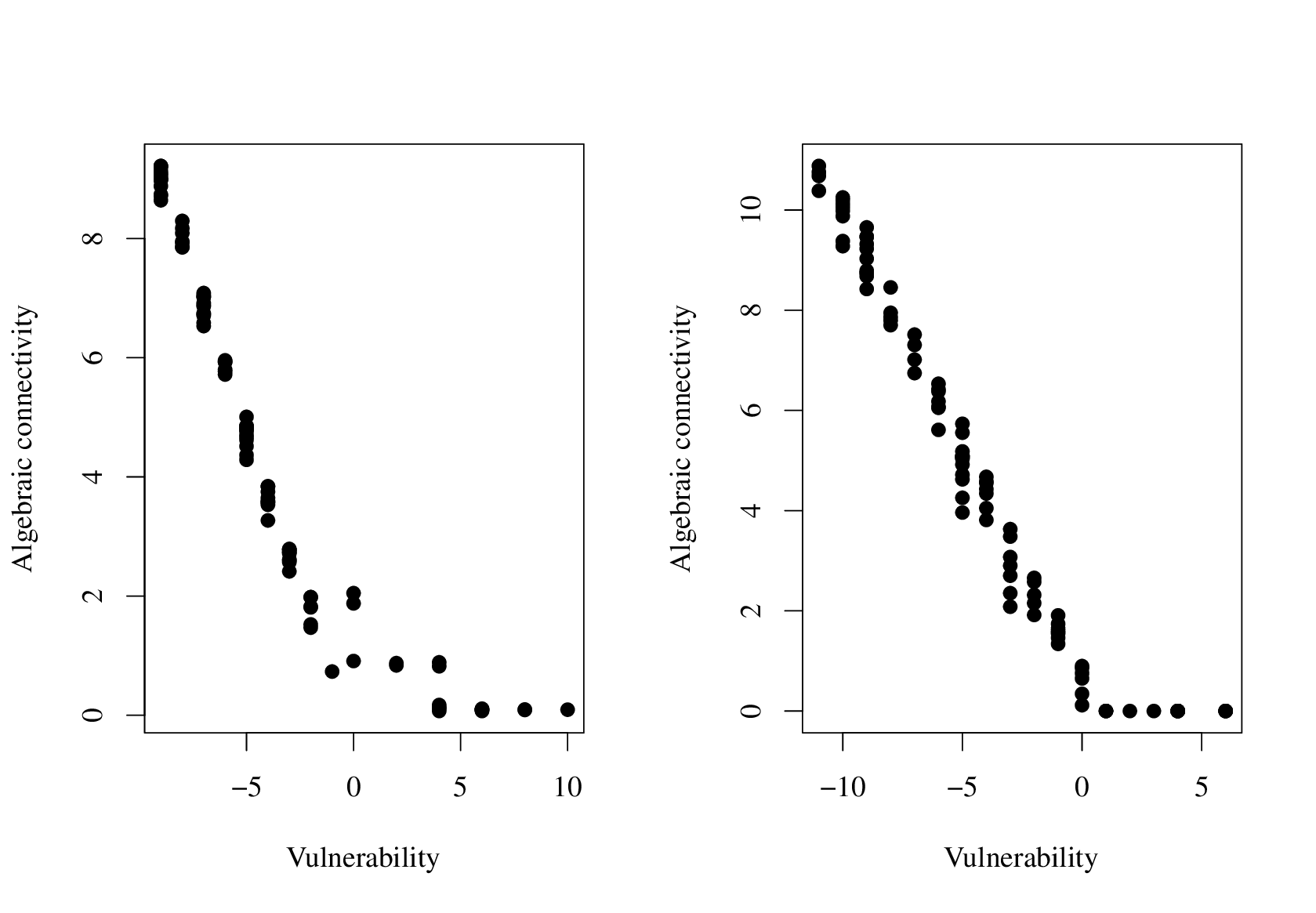}
\end{center}
\caption{Scatter plots comparing vulnerability and algebraic connectivity over Barab{\'a}si-Albert graphs (left plot) and Erd\H{o}s-R\'{e}nyi graphs (right plot).}
\label{vul_vs_ac}
\end{figure}

\subsection{Vulnerability and robustness} \label{robustness}

The goal of the first experiment is to assess the relationship  among vulnerability and robustness of a graph: are robust graphs less vulnerable? Do fragile networks have high vulnerability? 
For this experiment we generate random graphs according to the following two graph models: Barab{\'a}si-Albert graphs (BA graphs, for short), also known as scale-free graphs, and Erd\H{o}s-R\'{e}nyi graphs (ER graphs, for short). We first generate a sample of 100 random BA graphs, varying the edge density. In particular, we choose randomly the number of edges to add in each step of the preferential attachment process in the interval from $1$ to $n/2$, where $n$ is the number of graph nodes. Hence, both sparse and dense graphs are generated. Next, we generate a sample of the same size of random ER graphs according to the model $G(n,m)$; we generated the ER graphs with the same edge densities of the BA graphs previously sampled. On each graph of the sample, we compute the vulnerability and the algebraic connectivity. The \textit{algebraic connectivity} of a graph is the second-smallest eigenvalue of the Laplacian matrix of the graph. This eigenvalue is greater than 0 if and only if the graph is connected. The magnitude of this value reflects how easily a network can be divided: it is small for networks that can be easily partitioned in two groups of nodes, that is, the network divides by removing few edges from it, and it is large for networks that can be hardly partitioned in two fragments, that is, to divide the network a large number of edges must be removed. Algebraic connectivity is hence a measure of the robustness of networks \cite{N10}.

As shown in Figure \ref{vul_vs_ac}, for both BA and ER graphs, vulnerability and algebraic connectivity are negatively correlated as soon as vulnerability is lower than or equal to the watershed score of 0 (recall that the same score of vulnerability determines if the network is regularizable or not). This means that,  regularizable networks with low vulnerability have high algebraic connectivity, and hence are robust graphs. On the other hand, for graphs with positive vulnerability, that is, networks that are not regularizable, there is no association between vulnerability and algebraic connectivity.

Given these experimental outcomes, we conjecture a partial mathematical relationship between vulnerability and algebraic connectivity of networks.\footnote{This intuition is corroborated by the known result that expanders (see Section \ref{related}) are graphs with large algebraic connectivity.} 

A first step towards a precise formalization of this relationship is the following. Let $G=(V,E)$, with $|V|=n$ and let $S\subset V$. The set of the edges connecting $S$ with the rest of the graph makes up the boundary of $S$, that we denote with $\partial(S)$. Formally
$$\partial(S)=\{i j \in E: |S\cap\{i,j\}|=1\}.$$
Clearly, in the case where $S$ is an independent set then $$|\partial(S)|=\sum_{i\in S}|\partial(\{i\})|.$$ 
Actually, for every $S\subset V$ it turns out that $$\frac{|\partial(S)|}{|S|}\geq
\lambda_2 \Bigl(1-\frac{|S|}{n}\Bigr),$$ where $\lambda_2$ is the second-smallest eigenvalue of the graph Laplacian, that is, the graph algebraic connectivity \cite{CC01}. If $S$ is an independent set, then
$$\frac{|\partial(S)|}{|S|}=\frac{\sum_{i\in S}|\partial(\{i\})|}{|S|}$$ is the mean degree of the
nodes of $S$. For any node set $S$, we have that $|N(S)|$ is always greater than or equal to the maximum degree of the nodes in $S$, and hence, it is also greater than or equal to the mean degree of the nodes in $S$. Summing up, if $S$ is an independent set, we have $$\lambda_2
\Bigl(1-\frac{|S|}{n}\Bigr)\le \frac{|\partial(S)|}{|S|}     \le   |N(S)|.$$ This inequality is
weak and makes sense only for $\lambda_2>1$; however it partially explains the results of the
experiments: if algebraic connectivity ($\lambda_2$) is high, then, any independent set $S$ has a large set of neighbors $N(S)$, and hence the vulnerability of the graph cannot be large (see Figure \ref{vul_vs_ac}). 

Another simple observation helps us complementing the explanation of the experimental results. If a graph $G$ has two nodes of degree 1
connected to a third node (of arbitrary degree), then $1$ is an eigenvalue of the
Laplacian matrix \cite{CC01}, so that $\lambda_2\le 1$. But at the same time the graph vulnerability $ \vg_G \ge 1$, and, if the nodes of degree one connected to the same node are $k$, then $\vg_G \ge k-1$. This suggests that when algebraic connectivity is small ($\lambda_2 \le 1$) we cannot expect any relationship between vulnerability and algebraic connectivity (see again Figure \ref{vul_vs_ac}).

\begin{figure}[t]
\begin{center}
\includegraphics[scale=0.40, angle=-90]{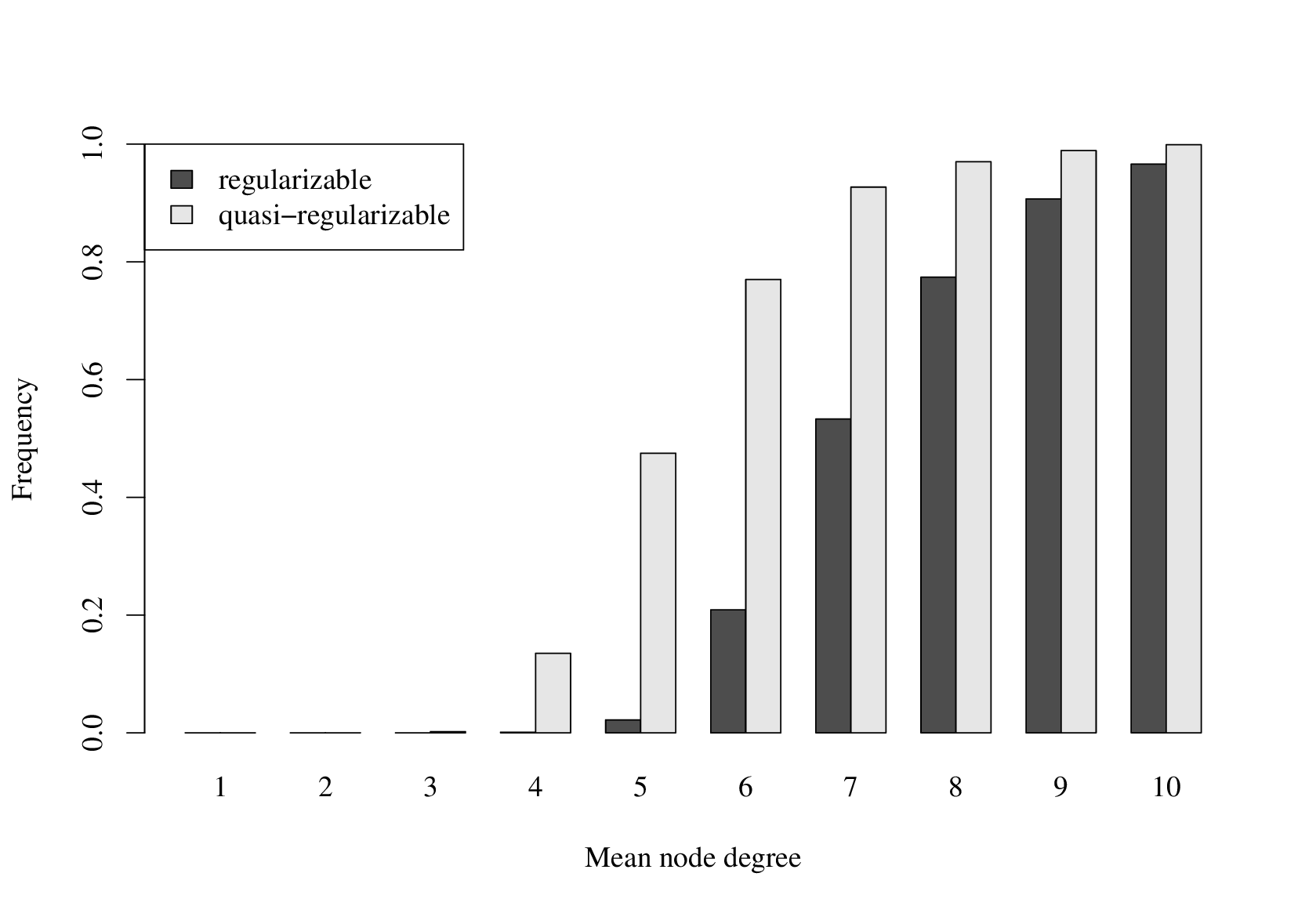}
\end{center}
\caption{Frequency of Erd\H{o}s-R\'{e}nyi graphs that are regularizable and quasi-regularizable by increasing the mean node degree.}
\label{reg-prob}
\end{figure}

\subsection{The frequency of vulnerable networks} \label{frequency}

The aim of the second experiment is to estimate the probability of being a regularizable or quasi-regularizable graph: how many graphs are regularizable? How many graphs are quasi-regularizable? Notice that, because of Theorem \ref{quasiregular}, a network is vulnerable if and only if it is not quasi-regularizable, hence the probability of finding a vulnerable network is the complement to 1 of the probability of finding a quasi-regularizable network. 

For this experiment, we generate a sample of Erd\H{o}s-R\'{e}nyi graphs, increasing the average node degree from 1 to 10. We use the model $G(n,p)$ of ER graphs, where $n$ is the number of nodes and $p$ is the probability of edges between vertices. The mean degree of a node in a $G(n,p)$ graph is $\langle k \rangle = p (n-1)$. We fix the number of nodes $n = 100$ and increase $p$ so that we obtain the mean degree sequence from $1$ to $10$. For each pair $(n,p)$, we generate a sample of 100 graphs according to the model $G(n,p)$ of ER graphs.
For each graph in the sample, we check whether the graph is regularizable and, if not, whether it is quasi-regularizable. As it is clear from Figure \ref{reg-prob}, the frequency of quasi-regularizable graphs and that of regularizable graphs increase as the mean node degree $\langle k \rangle$ grows. More precisely, when $\langle k \rangle$ is low, both frequencies are negligible. As soon as $\langle k \rangle$ is sufficiently large, both frequencies start growing very rapidly. By way of example, when $n = 100$,  the frequency of quasi-regularizable graphs is negligible as soon as $\langle k \rangle \leq 3$, it is significantly above 0 (14\%) when $\langle k \rangle = 4$,  when $\langle k \rangle = 5$ almost half (48\%) of the graphs in the sample are quasi-regularizable, and as soon as $\langle k \rangle = 6$ more than three-quarters (77\%) of the sampled random networks are quasi-regularizable. For higher values of the mean node degree, the frequency of quasi-regularizable graphs is close to 100\%.  As for regularizability, the frequency is negligible as soon as $\langle k \rangle \leq 5$. Graphs with $\langle k \rangle = 6$ have 21\% probability of being regularizable, those with $\langle k \rangle = 7$ have 50\% chance of being regularizable, while networks with $\langle k \rangle \geq 9$ are almost certainly regularizable. We notice, however, that these frequencies tend to become lower as soon as the number of nodes increases. 

We conjecture that there exists a transition phase of regularizability of networks that depends predominantly on the mean degree of the network.\footnote{A similar transition phase has been noticed for the giant component of networks: as soon as the mean degree of a node is higher than 1, a giant connected component including the majority of the graph nodes emerges \cite{N10}.} This seems reasonable with the benefit of hindsight. Recall that regularizability is the process of assigning weights to edges so that the resulting graph is regular. When the mean node degree is low, nodes have few incident edges, hence the process of regularizability is hampered. However, as soon as node degrees grow, there are many more possibilities of assigning weights to edges, significantly increasing the probability of success of the regularizability process. Finally, when node degrees are sufficiently large, there are so many possible weight assignments that the graph is almost certainly regularizable.

\subsection{Vulnerability and power on real networks} \label{real}

\begin{table}
\begin{center}
\begin{tabular*}{1\textwidth}{@{\extracolsep{\fill}}lrrrrrrr}
\textbf{network} &  \textbf{nodes} & \textbf{edges} & \textbf{vul} & \textbf{maxdeg} & \textbf{maxpow} & \textbf{maxdiff} & \textbf{cor} \\ \hline
madrid & 64 & 243 & 1 & 29 & 2.89 & 0.54 & 0.84  \\ \hline
netsci & 379 & 914 & 14 & 34 & 8.85 & 0.49 & 0.89  \\ \hline
powergrid & 4941 & 6594 & 575 & 19 & 9.73 & 0.73 & 0.84  \\ \hline
internet & 22963 & 48436 & 16362 & 2390 & 1127.77 & 0.05 & 0.97  \\ \hline
\end{tabular*}
\end{center}
\caption{Statistics for the four analyzed networks. The meaning of columns is: network: name of the network; nodes: number of nodes; edges: number of edges; vul: vulnerability; maxdeg: maximum degree of a node; maxpow: maximum power of a node; maxdiff: maximum difference in power among nodes with the same degree divided by the maximum difference in power among any two nodes (runs between 0 and 1); cor: Pearson correlation coefficient between degree and power (runs between -1 and 1).}
\label{statistics}
\end{table}

In our last experiment we apply the developed vulnerability and power measures to real-world networks. The goal of this experiment is twofold: (i) show that  vulnerability and power measures might reveal meaningful properties of the structure of a network; (ii) empirically study the correlation among Shapley-based node power\footnote{In this section we use power defined as $\phi^{p}$ in Theorem \ref{th:power} of Section \ref{gamepower}.} and node degree in a network. We analyzed four real networks, two social networks and two technological networks. Table \ref{statistics} summarizes some statistics we have computed on these networks.

\begin{figure}
\begin{center}
\includegraphics[scale=0.40, angle=-90]{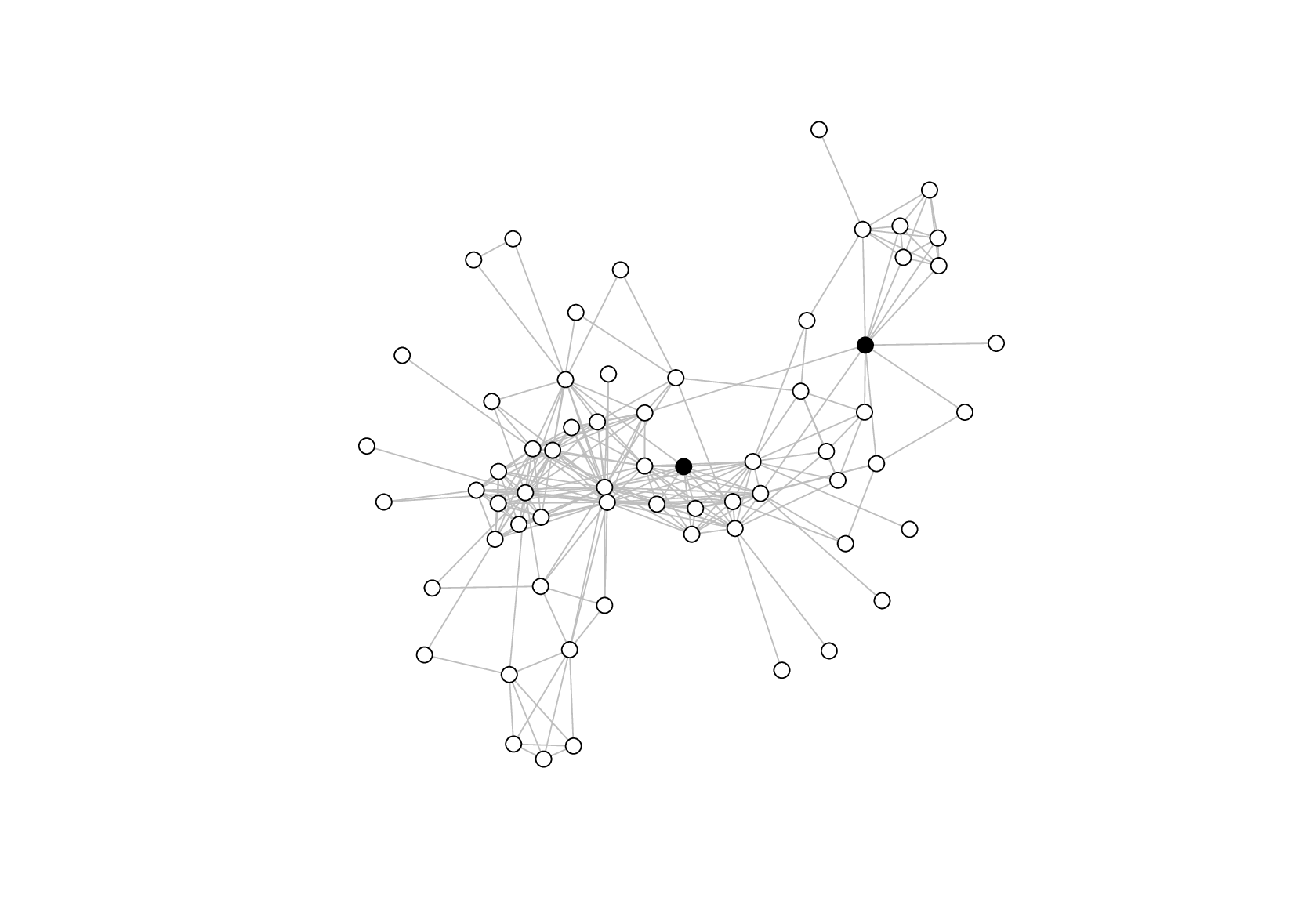}
\end{center}
\caption{Madrid train bombing terrorist network. Black circles are, among nodes having the same degree, those having maximum power difference (54\% of the size of the power range).}
\label{madrid}
\end{figure}

\begin{figure}
\begin{center}
\includegraphics[scale=0.40, angle=-90]{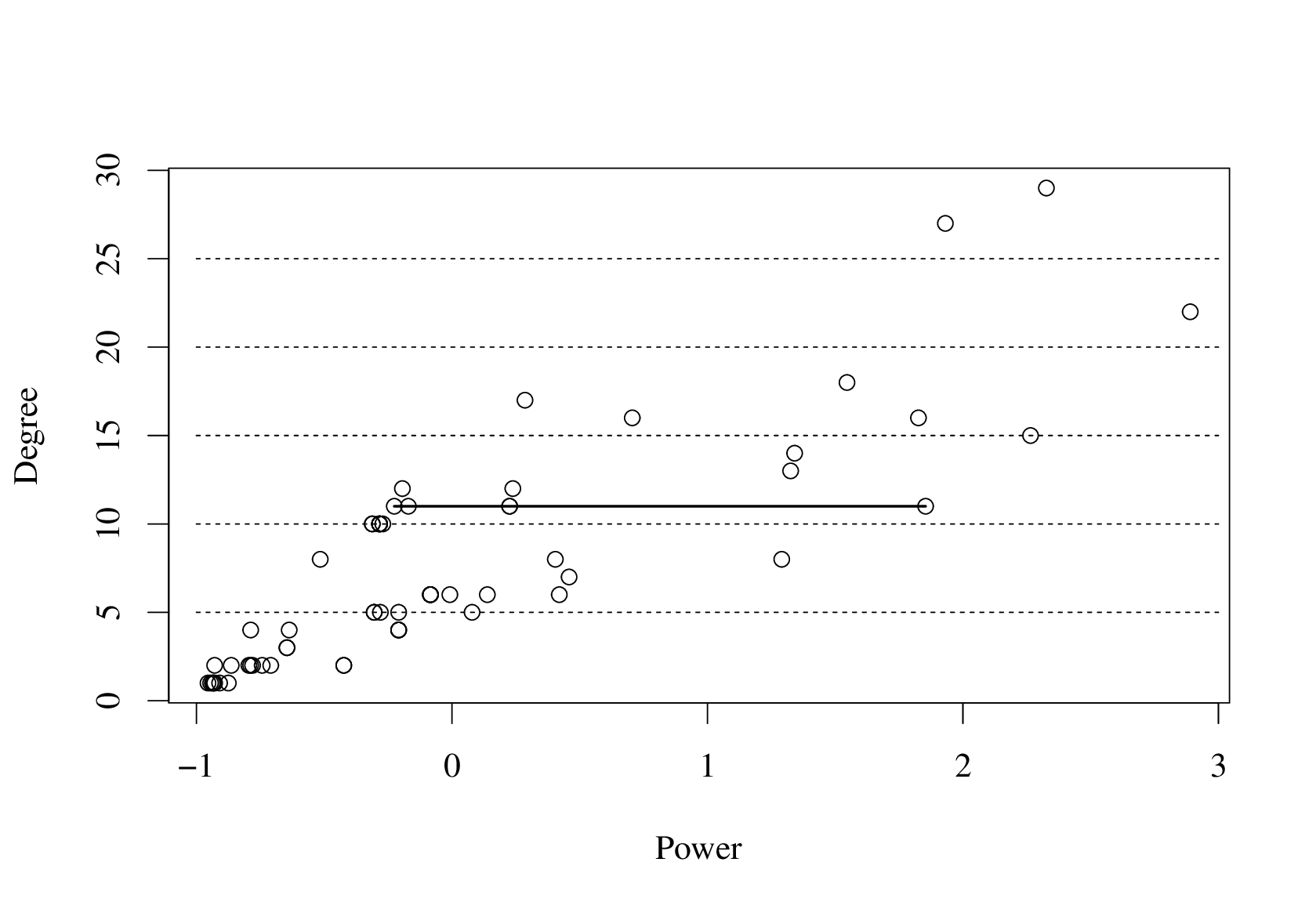}
\end{center}
\caption{Scatterplot between power and degree of nodes of the Madrid train bombing terrorist network. The extreme circles connected by the horizontal segment are, among nodes having the same degree, those having maximum power difference (54\% of the size of the power range).}
\label{madrid-scatter}
\end{figure}

The first social network is the Madrid train bombing terrorist network. The network depicts individuals involved in the bombing of commuter trains in Madrid on March 11, 2004. Ties link the individuals involved in at least one of the following relationships: (1) trust or friendship; (2) ties to Al Qaeda and to Osama Bin Laden; (3) co-participation in training camps or wars; (4) co-participation in previous terrorist attacks. The network was reconstructed by Jos\'{e} A. Rodr\'{i}guez of the University of Barcelona using press accounts in the two major Spanish daily newspapers \cite{H06}. It is depicted in Figure \ref{madrid}. 

The vulnerability score of the terrorist network is very low. In fact, as soon as one removes the nodes with degree equal to 1, the resulting network becomes regularizable, with a negative vulnerability score equal to -1. Also, there are no big differences among the power scores of nodes: the great majority of the terrorists (84\%) have power between -1 and 1, with a maximum power of 2.89. If follows that the terrorist network contains no core-periphery, executioner-victims fragment, in which an independent group of terrorists is connected to a unique central control. On the contrary, the network is composed of few communities, one of them quite prominent, of tightly connected individuals, with few links among the different communities \cite{H06}. This flattened, non-hierarchical, and decentralized layout, with no leader in control and defined ranks, is a form of robustness against attacks: no individual is fundamental for the network, and when some terrorist is removed (jailed, for instance), new substitutes immediately emerge.   

The correlation among degree and power is depicted in the scatterplot of Figure \ref{madrid-scatter}. Although there exists a positive correlation among the two measures (the Pearson correlation coefficient is 0.84), degree alone cannot explain power. Indeed, there are nodes with similar degree having quite different power, so that the points in the plot do not follow a straight line but are dispersed in a fan-like shape. Both the scatterplot and the network figures highlight the node pair with same degree and maximum power divergence. Despite this two nodes have the same degree (11), it is clear from the network visualization that they have different structural roles: the less powerful individual is central to a big clique, and is surrounded by highly connected neighbors (on average its neighbors have degree 16), while the other one is a broker between scarcely connected neighbors (with an average degree of 6).    

\begin{figure}
\begin{center}
\includegraphics[scale=0.45, angle=-90]{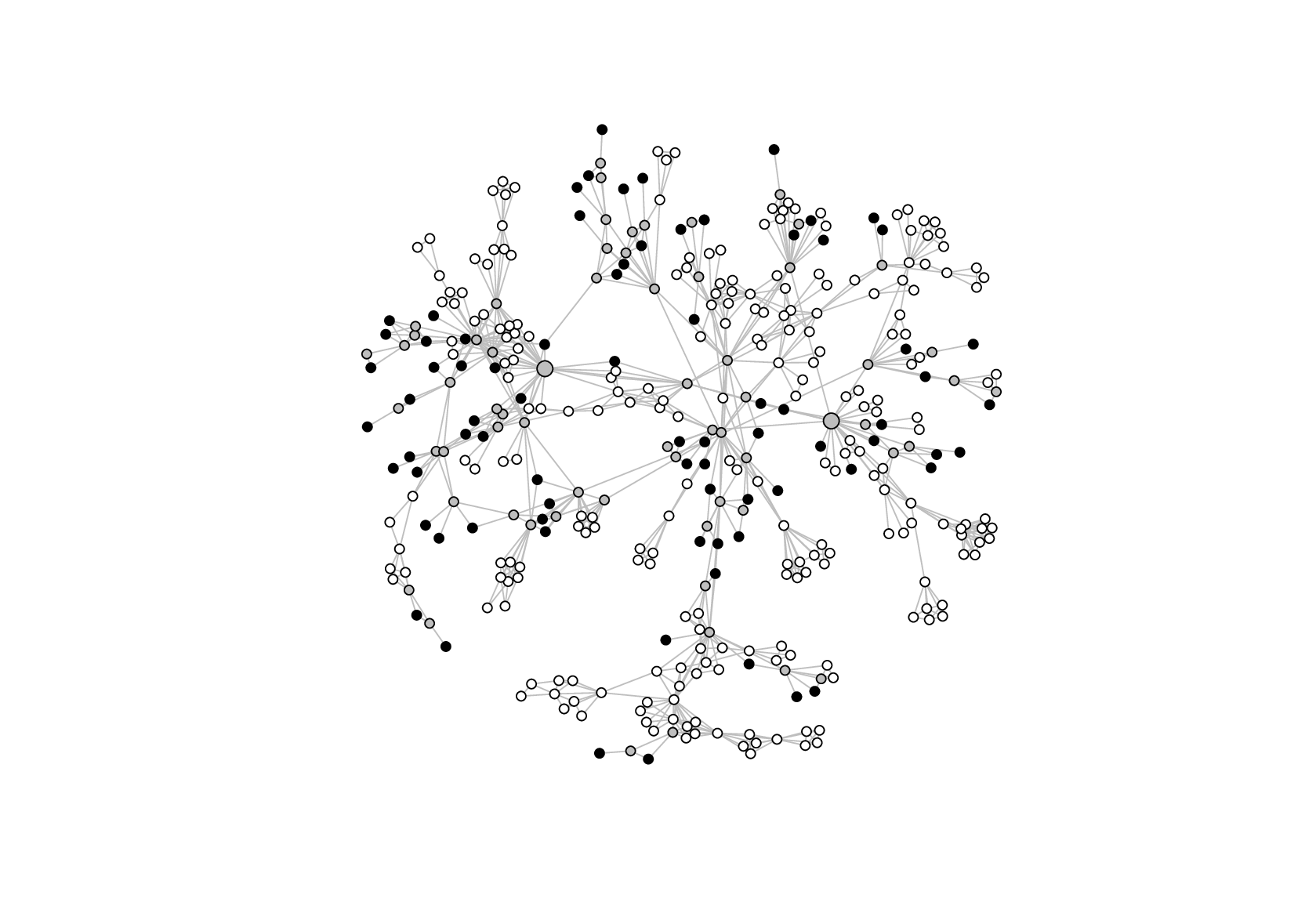}
\end{center}
\caption{Network science collaboration network. Black nodes form an independent set of maximum vulnerability (14): it contains 78 nodes and is dominated by the set of 64 grey nodes. The two bigger grey nodes have the same degree (27) and, among nodes having the same degree, they have the maximum power difference (49\% of the size of the power range): they are Hawoong Jeong (on the left), and Mark Newman (on the right). They are highlighted in Figure \ref{netsci-ego}.}
\label{netsci}
\end{figure}

\begin{figure}
\begin{center}
\includegraphics[scale=0.22, angle=-90]{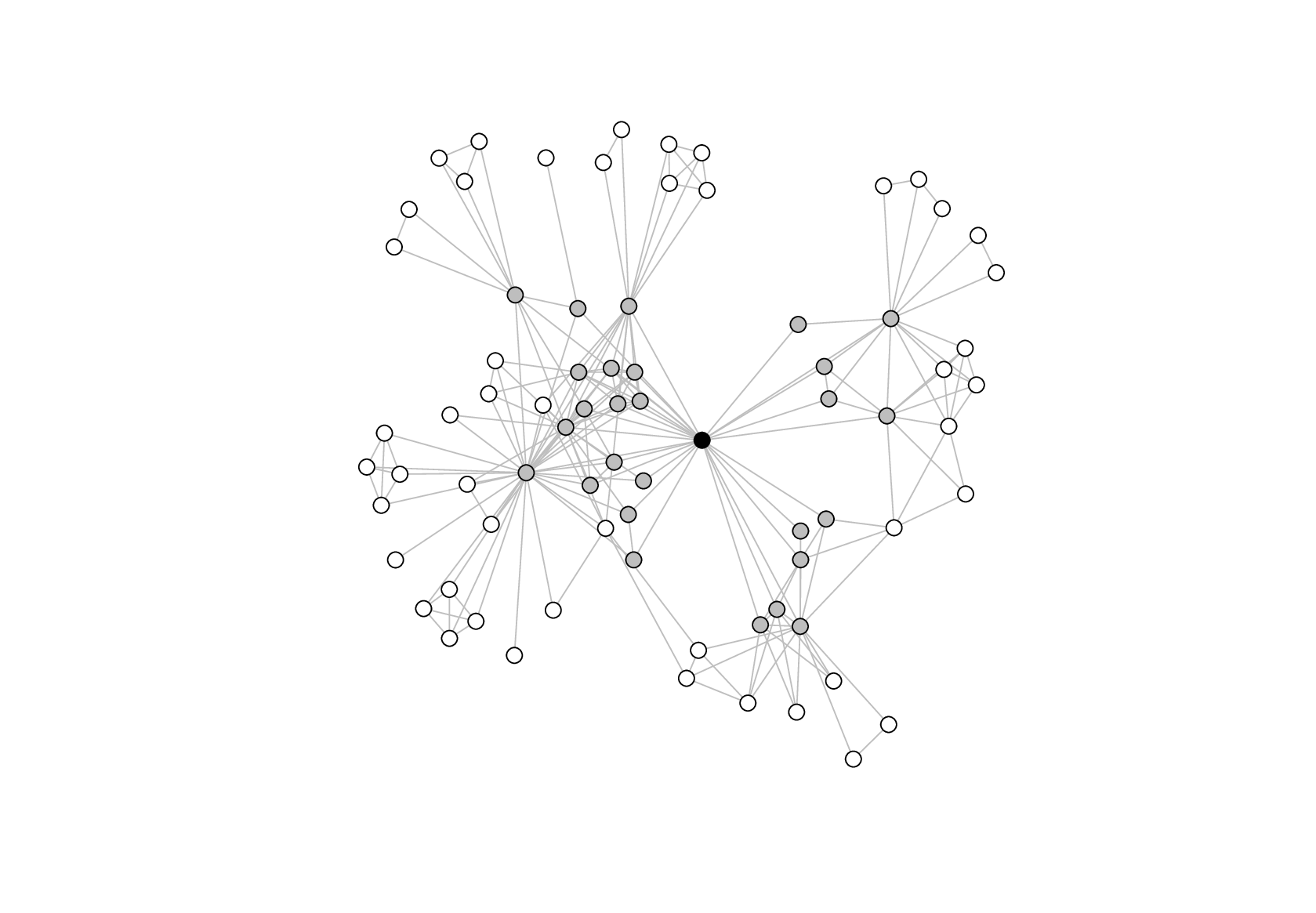}
\includegraphics[scale=0.22, angle=-90]{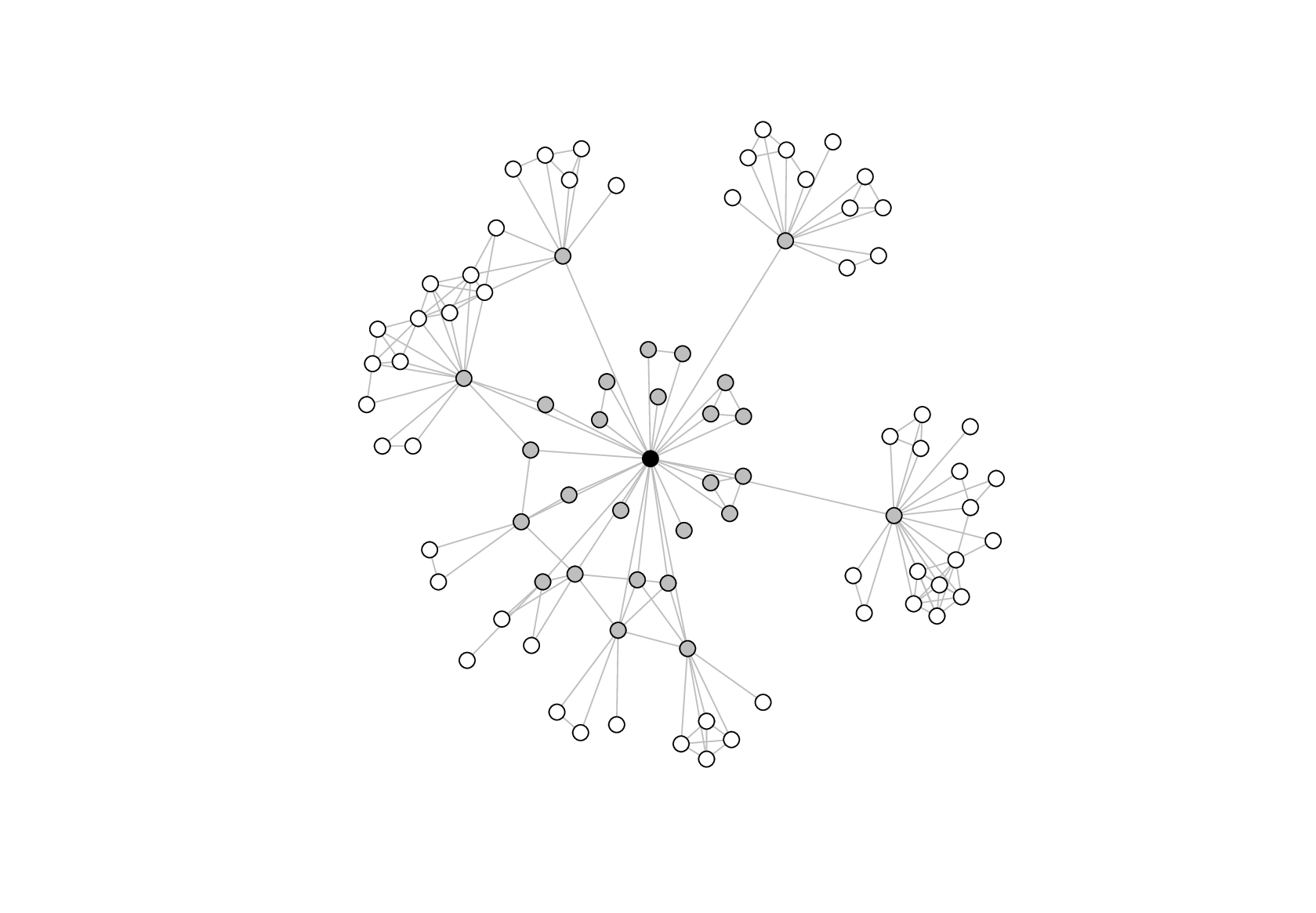}
\end{center}
\caption{The ego-centered networks of Hawoong Jeong (on the left), and Mark Newman (on the right). They depict the ego (black), their collaborators (grey), and the collaborators of their collaborators (white).}
\label{netsci-ego}
\end{figure}

The next network we analyze is a collaboration network of scholars in the field of network science. The nodes are scientists working on network theory and experiment, as compiled by Mark Newman in May 2006 \cite{NG04}, using the bibliographies of two main review articles on networks. There is a link between two authors if they have collaborated in at least one paper. The original version contains all
components of the network, for a total of 1589 scientists; here we study the
largest component of 379 scientists, which is depicted in Figure \ref{netsci}. 

With respect to the terrorist network, the collaboration network has a higher vulnerability (14 versus 1) and, although the largest degree of a node in the two networks is comparable (34 versus 29), the power spans a much larger interval (8.85 versus 2.89). This means that the structure of the network is more star-like, with core scholars that attract collaborators with a much fewer collaboration degree.  For instance, the most powerful scholar is Mark Newman (the bigger grey node on the right in Figure \ref{netsci}), with power 8.85. He has 27 collaborators, who are much less collaborative (their average degree is less than 5).

Again, we noticed a positive correlation between power and degree (Person correlation coefficient 0.89), but important divergences exists. For instance, the two scholars with the same degree and the maximum divergence in power are Hawoong Jeong (degree: 27, power: 4.02), and Mark Newman (degree: 27, power: 8.85), with a difference in power that accounts almost half of the power range. Their ego-centered sub-networks are depicted in Figure \ref{netsci-ego}. Notice that Jeong has more collaborative co-authors than Newman (the average collaboration degree is 8.4 for Jeong and 4.9 for Newman).

\begin{figure}
\begin{center}
\includegraphics[scale=0.45, angle=-90]{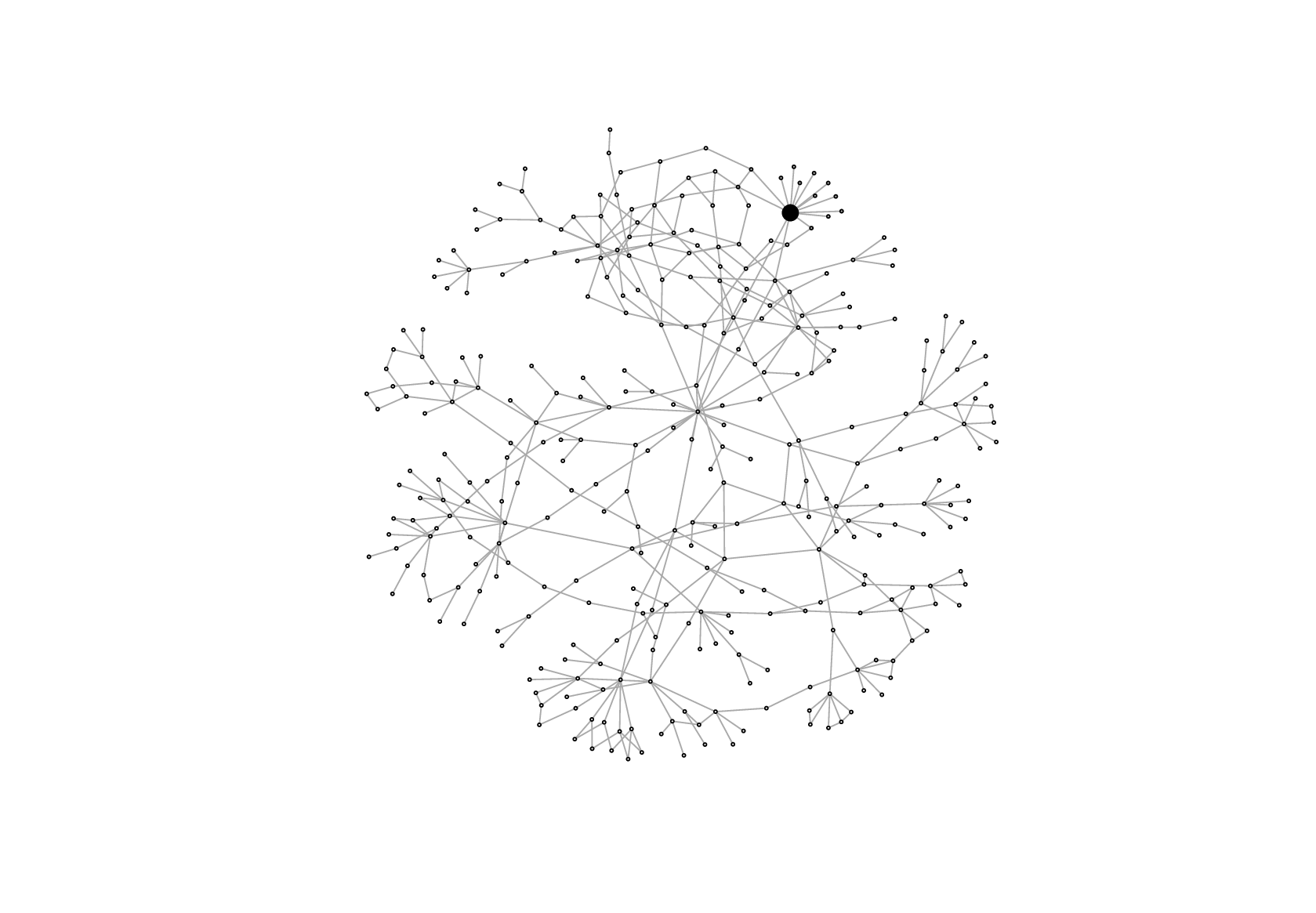}
\end{center}
\caption{A snapshot of the power grid network. It is the ego network of order 8 (containing all nodes at a distance less than or equal to 8 from the ego) centered at the node with maximum power (the bigger node).}
\label{power}
\end{figure}

\begin{figure}
\begin{center}
\includegraphics[scale=0.40, angle=-90]{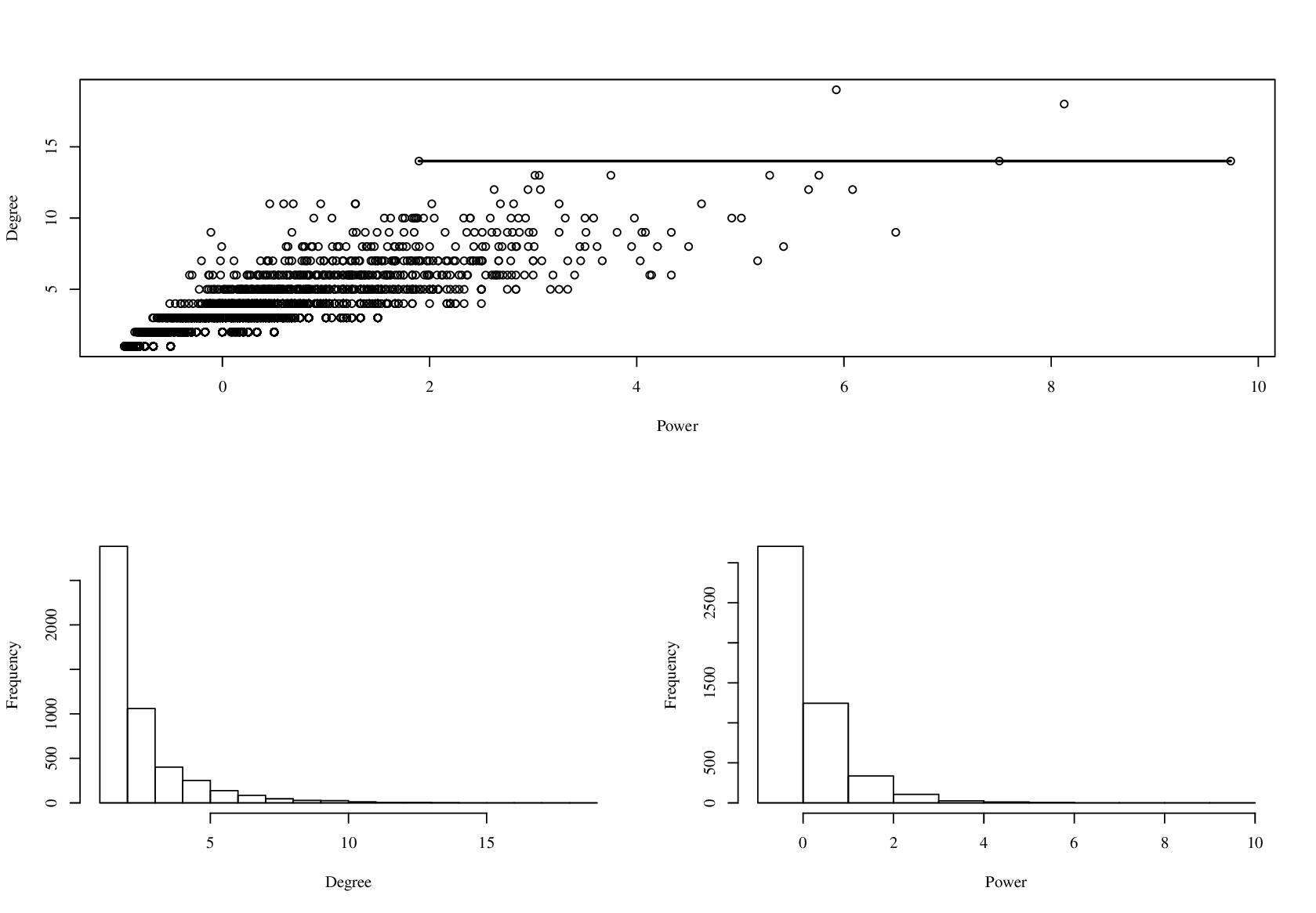}
\end{center}
\caption{Scatterplot between power and degree of nodes of the power grid network (above). The extreme circles connected by the horizontal segment are, among nodes with the same degree, those having maximum power difference (73\% of the size of the power range). Histograms of degree and power are shown below.}
\label{power-scatter}
\end{figure}

The last two graphs we investigate are two technological networks. The first is a representation of the topology of the western states power grid of the United States, compiled by Duncan Watts and Steven Strogatz \cite{WS98}.  The nodes are the generating stations and switching substations while the edges are the physical electric lines connecting them. A fragment of the network, which is much larger than the previously analyzed social networks, is depicted in Figure \ref{power}. 

The nodes of the power network have a relatively low degree: the typical station has two or three connections with other stations, while few hub stations have a larger number of connections, with a maximum degree of 19. The distribution of node power is similar, with the great majority of nodes with low power and a few of them with moderately high power, with a maximum of 9.73. The histograms of degree and power are depicted in the lower part of Figure \ref{power-scatter}. Degree and power are positively associated (Pearson 0.84), but, as clear from the scatterplot of the upper part of Figure \ref{power-scatter}, there are nodes with similar power and quite different degrees and nodes with similar degree and quite diverging power. This produces a scatterplot with a wide and high cloud of points (as opposed to a straight thin line).

Nevertheless, the vulnerability of the power network is significantly high: 575, more than 11\% of the number of nodes. There exists, indeed, an independent set of size 2264 that is dominated by a set size 1689. Such a high network vulnerability, with a relatively modest power at the level of nodes, reveals the particular network topology of the power grid network. Nodes are mostly arranged along linear paths. This is because edges represent physical lines, which, for economical reasons, typically connect geographically close stations. Hence, it is likely that two far away stations are connected through a chain of intermediated linked stations. Moreover, some stations are more important than others, and are connected to a moderate number of other independent stations, in a star-like structure. The resulting topology has large tree-like fragments, although the overall network contains circuits, as evident from the visualization offered in Figure \ref{power}.  

\begin{figure}
\begin{center}
\includegraphics[scale=0.45, angle=-90]{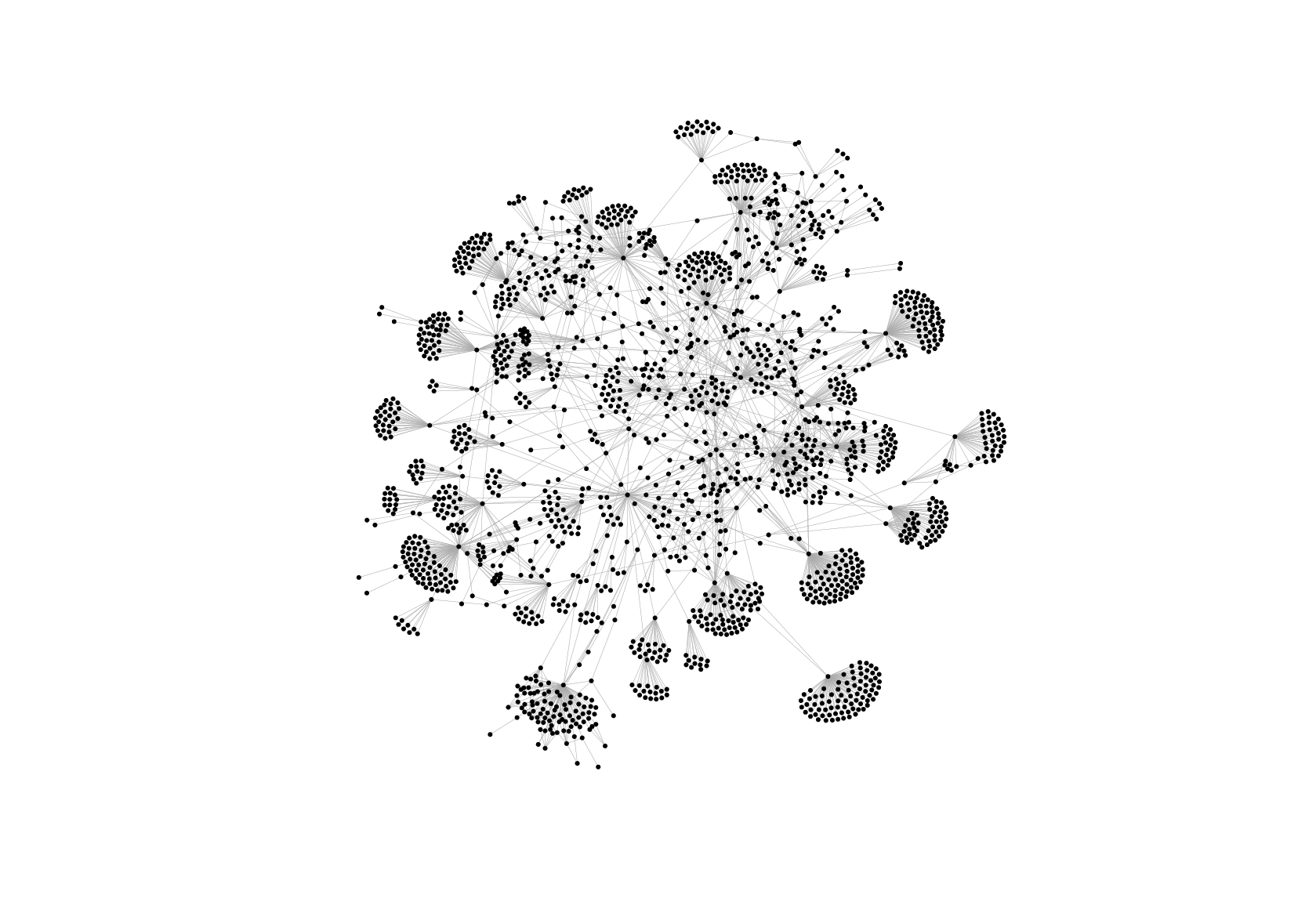}
\end{center}
\caption{A fragment of the Internet consisting of an ego network of order 4 centered at the node of maximal power. For the sake of visualization, only nodes with maximum degree 100 are considered.}
\label{internet}
\end{figure}

The last network we observe is the technological network by definition: the Internet. The representation we use contains a symmetrized snapshot of the structure of the Internet at the level of autonomous systems, reconstructed from Border Gateway Protocol tables posted at \url{archive.routeviews.org}. Nodes represent autonomous systems -- collections of computers and routers, usually under single administrative control, within which data routing is handled independently of the wider Internet. Edges are physical data connections between these systems. This snapshot was created by Mark Newman from data for July 22, 2006.

It is immediately clear from the figures in Table \ref{statistics} that this network is different from the previous ones. The distributions of degree and power are severely skewed, with relatively few hub systems that draw the majority of connections. For instance, 75\% of the systems have one or two connections, 95\% have less than 9 connections, and 99\% have less than 37 connections. There are 76 hubs with more than 100 connections, 6 of them have more than 1000 connections, and the most linked node has 2390 connections, reaching 10\% of the graph. The high asymmetry determines a high Pearson correlation coefficient among degree and power (0.97) and a low maximum power divergence among same-degree nodes (0.05). However, these figures are artifacts of the huge skewness of the distributions of power and degree. Indeed, the (non-parametric) Spearman rank correlation coefficient between degree and power is much lower: 0.48. This means that, also for the Internet, degree only partially explains power of a node. 

The vulnerability of the network is extremely large: there exists an independent set of cardinality 19018 (notably, 83\% of the network) that is dominated by a much smaller set of 2656 nodes, making the vulnerability of the network equal to the whopping 16362. These figures reveal a network dominated by few powerful hubs. This core, made of high-performance routers and long-distance high-bandwidth lines, is well known as the backbone of the Internet. It provides connection to a plurality of Internet Service Providers, who in turn  serve connectivity to a myriad of end users, the ultimate consumers of the Internet bandwidth. This peculiar topology, illustrated in Figure \ref{internet}, is also responsible for the vulnerability of Internet to attacks. Since there is so much control in relatively few hubs, a malicious individual can take advantage of this topology flaw by attacking few crucial routers and causing conspicuous effects.   

\section{Related literature} \label{related}

The notion of vulnerability we have proposed is somewhat related to that of expander graph \cite{HLW06}. Informally, an expander graph is an undirected unweighted graph that is both sparse and robustly connected. Sparsity is achieved by constraining all nodes of the graph to have the same small degree $k$, which is constant with respect to the number of nodes $n$ (hence expander graphs are $k$-regular graphs). Robustness holds since every not too large subset of nodes of an expander graph has a relatively large boundary, where the boundary $\partial S$ of a node set $S$ is defined as the set of edges emanating from $S$ to its complement. The expansion parameter for a regular graph $G$ is defined as $$h(G) = \min_{S : |S| \leq n/2} \frac{|\partial S|}{|S|}$$ and a regular graph is a good expander if its expansion parameter is well above $0$. 

Expanders can be defined and investigated in different languages including graph theory, geometry, probability and algebra. In graph theory, expanders are graphs that are both sparse (hence economical) and robust (to failure or attacks):  to disconnect a large part of the graph, one has to remove many edges. Using the geometric notion of isoperimetry, every set of vertices of an expander graph has a relatively large boundary. From the probabilistic perspective, expanders are graphs for which a natural random walk on the graph converges to its limiting distribution very rapidly. Algebraically, expanders are graphs with a large eigengap between the largest and second-largest eigenvalues of the adjacent matrix of the graph (this property is related to the convergence speed of the above mentioned random walk on the graph). Equivalently, expanders are graphs with a large second-smallest eigenvalue of the Laplacian matrix of the graph (algebraic connectivity), and hence are robust graphs.     

Recall that we defined vulnerability of an arbitrary graph as $$\vg_G = \max_{\emptyset \neq S \in {\cal S}(G)} |S| - |N(S)|.$$ Our definition diverges from that of expander graph for the following reasons:

\begin{enumerate}
\item expansion is a bound on the ratio between a number of edges and a number of vertices, whereas vulnerability takes the difference between two sets of vertices. This is a huge gap -- for instance, the boundary of the set of leaves in the star graph with $n$ nodes has size $n - 1$, whereas the size of the neighbor set of the leaves is 1; 

\item vulnerability is defined on arbitrary graphs, while an expander is a $k$-regular graph with small $k$;

\item finally, in the context of network science, graph expanders have been studied with the goal of \textit{designing} future communication networks with good topological properties, while we propose graph vulnerability with the aim of \textit{analyzing} existing real networks.
\end{enumerate}

The Shapley value-based node power introduced in this paper is also weakly related to the sociological theory of structural holes  \cite{B04}. The author argues very convincingly that ``opinion and behavior are more homogeneous within than between groups, so people connected across groups are more familiar with alternative ways of thinking and behaving. Brokerage across the structural holes between groups provides a vision of options otherwise unseen, which is the mechanism by which brokerage becomes social capital. [...] Compensation, positive performance evaluations, promotions, and good ideas are disproportionately in the hands of people whose networks span structural holes''. In short, these social brokers ``see bridges where others see holes''. A quantitative measure of the mentioned local betweenness centrality is the local clustering coefficient \cite{WS98,N10}. For a given node $i$, the local clustering coefficient is the ratio of the number of pairs of neighbors of $i$ that are connected and the number of pairs of neighbors of $i$. This coefficient is low if there are many structural holes among the neighbors of node $i$, making the subgraph induced by the neighborhood of $i$ loosely connected. In such a case the broker $i$ has power over information flow between those friends that are not directly connected. The coefficient is high if the neighbors of $i$ are instead tightly connected, and information between these friends can flow directly without passing through $i$, lowering the power of $i$. In fact, the \textit{inverse} of the local clustering coefficient might be regarded as a centrality measure of \textit{local betweenness} \cite{N10}.

Now consider a powerful node. Since, by definition of power, the node has many neighbors with low degree, we might expect that the node has low clustering coefficient, hence high local betweenness. However, a node $i$ with high local betweenness is not necessarily a powerful node, since the set of neighbors of $i$ might be well connected to nodes different from $i$, and hence $i$ might be powerless. 

Standard node centrality measures, like degree, closeness and betweenness, have been extended to sets of nodes \cite{EB99}. In particular the authors define group degree centrality as the relative number of non-group nodes that are connected to group members, that is, for a node set $S$ in a graph with nodes in $V$, group degree centrality is $$\delta(S) = \frac{|N(S) \setminus S|}{|V \setminus S|}.$$ The coefficient runs from 0 to 1 and, assuming a connected graph, it is maximum for \textit{dominating sets} $S$ such that every node not in $S$ is adjacent to at least one member of $S$. To be effective, it would be desirable for the group $S$ to be as small as possible without sacrificing centrality \cite{EB99}. Therefore, the authors propose to search for the smallest set $S$ with the maximum degree centrality, that is, the smallest dominating set. In graph theory, the cardinality of the smallest dominating set is known as \textit{domination number} of the graph, and finding the domination number of an arbitrary graph is a classical computationally hard problem. Therefore it is believed that there is no efficient algorithm that finds a smallest dominating set for a given graph.
The problem of finding the smallest dominating set bears some analogy with that of finding the set of maximum power in our setting. However, there are also significant differences: while the former problem searches for a small set with a neighbor set that expands over the whole graph, the latter seeks for a small set that controls a large (independent) set. 

The first application of game theory to the topic of network centrality used the  Banzhaf power index instead of the Shapley value \cite{GO82}. The use of the Shapley value as a network centrality measure has been later investigated  \cite{SN10,MASRJ13,SMR12}. The authors consider the node-set generalizations of the principal centrality measures, including degree, closeness, and betweenness, and interpret them as characteristic functions of coalitional games. Then, the Shapley value of these games is proposed as a more involved centrality index at node level. Moreover, polynomial time solutions for Shapley value-based degree, closeness, and betweenness centrality have been devised \cite{MASRJ13,SMR12}. We follow a similar technique to introduce closed-form polynomial-time expressions for the Shapley value of vulnerability and power measures.

\section{Conclusion} \label{conclusion}

We have defined a vulnerability measure on sets of nodes of a network that counts the difference between the number of nodes in the set and the number of neighbors of nodes in the set. The measure is seemingly simple, but has proved interesting from a theoretical, computational and empirical point of view. 

We have thoroughly investigated the problem of finding a non-empty independent set of maximum vulnerability in a graph. The vulnerability of a graph, defined as the optimal value for the problem, provides a partition of the class of networks into regularizable graphs (those with negative vulnerability), quasi-regularizable graphs that are not regularizable (those with null vulnerability), and graphs that are not quasi-regularizable (those with positive vulnerability). 

Computationally, the maximum vulnerability problem can be solved efficiently, by reducing to the minimum 2-vertex cover problem, for the class of non-regularizable graphs (those with null or positive vulnerability). The complexity is $O(|V|^{\frac{1}{2}} \cdot E)$ for graphs with positive vulnerability, and $O(|V|^{\frac{3}{2}} \cdot E)$ for graphs with null vulnerability. These bounds boil down to $O(|V|^{\frac{3}{2}})$ and $O(|V|^{\frac{5}{2}})$ on sparse networks with $m = O(n)$. 
Furthermore, we have modelled the maximum vulnerability problem in integer linear programming, showing that a single continuos relaxation of the model is sufficient to solve the problem on non-regularizable graphs, while, for regularizable networks, the solution of $|V|$ linear programming instances are necessary. Incidentally, this demonstrates that the maximum vulnerability problem is polynomial and provides a practical, highly efficient and optimized method (linear programming) to tackle the problem. 

We have interpreted the vulnerability measure (as well as its mirror image power measure) as the characteristic function of a coalition game played on the graph and have proposed the Shapley value of the game as a sophisticated measure of vulnerability (and power) at the level of nodes. Interestingly, the emerging measure of power pontificates that power is in the hands of those connected to powerless ones, a thesis that was already suggested in the sociological literature of the late sixties. Moreover, the measure has a closed-form expression that can be computed in linear time in the size of the graph. 

We have experimentally shown on artificial graphs (using both random and scale-free models) that a network is almost certainly non-regularizable when its mean node degree is sufficiently small. Hence, sparse networks tend to be non-regularizable. This is good news, since most real networks are sparse -- we have analyzed two social networks and two technological networks (including the Internet) and found that they are, indeed, non-regularizable. This opens the possibility of applying the developed measures, at both group level and node level, to large real networks.

\bibliographystyle{plain}

\end{document}